\documentclass[11pt,a4paper]{article}

\usepackage{bold-extra}

\usepackage{graphicx}

\graphicspath{ {./Figures/} }

\usepackage{comment}

\pdfoutput=1
\usepackage{color}
\usepackage{todonotes}
\usepackage{subfigure}
\usepackage{booktabs}
\usepackage{array}
\usepackage{multirow,rotating}
\usepackage{amsmath,amssymb,bm,cite,graphicx,wasysym,mathrsfs,dsfont}
\usepackage{slashed,relsize}
\usepackage[colorlinks=true,linkcolor=black,anchorcolor=black,citecolor=blue,filecolor=cyan,menucolor=red,runcolor=filecolor,urlcolor=blue,bookmarks=true,bookmarksnumbered=true]{hyperref}

\usepackage[font=small,labelfont=bf]{caption}

\def\beq{\begin{equation}\displaystyle\displaystyle}
\def\eeq{\end{equation}}
\def\bea{\begin{eqnarray}\displaystyle} 
\def\eea{\end{eqnarray}}

\def\({\left(}
\def\){\right)}
\def\bry{\begin{array}}
\def\ery{\end{array}}

\title{
\vspace{-2cm}
\vspace{3cm}
\bf \LARGE
Bubble wall dynamics\\ at the electroweak phase transition
\vspace{.2cm}}
\date{}
\author{
{\large Stefania De Curtis$^a$, Luigi Delle Rose$^b$, Andrea Guiggiani$^a$, \'Angel Gil Muyor$^b$,}\protect\\
{\large and Giuliano Panico$^a$}\\
[7mm]
\normalsize\itshape $^a$ INFN Sezione di Firenze and Dipartimento di Fisica e Astronomia, \protect\\ Universit\`a di Firenze, Via G. Sansone 1, I-50019 Sesto Fiorentino, Italy\\
\normalsize\itshape $^b$ IFAE and BIST, Universitat Aut\`onoma de Barcelona, 08193~Bellaterra,~Barcelona,~Spain
}

\begin{document}
\baselineskip=14pt
\arraycolsep=2pt

\begin{flushright}
$ $
\end{flushright}

\vspace{2em}

{\let\newpage\relax\maketitle}
\begin{abstract}
\medskip
\noindent
First order phase transitions could play a major role in the early universe, providing important phenomenological
consequences, such as the production of gravitational waves and the generation of baryon asymmetry.
An important aspect that determines the properties of the phase transition is the dynamics of the true-vacuum bubbles,
which is controlled by the density perturbations in the hot plasma. We study this aspect presenting, for the first time, the full solution
of the linearized Boltzmann equation for the top quark species coupled to the Higgs field during a first-order electroweak phase transition. Our approach, differently from the traditional one based on the fluid approximation, does not rely on any ansatz and can
fully capture the density perturbations in the plasma. We find that our results significantly differ from the ones obtained in the
fluid approximation (including its extensions and modifications), both at the qualitative and quantitative level. In particular
sizable differences are found for the friction acting on the bubble wall.
\end{abstract}

\vfill
\noindent\line(1,0){188}\\
{\scriptsize{E-mail: \texttt{stefania.decurtis@fi.infn.it}, \texttt{ldellerose@ifae.es}, \texttt{andrea.guiggiani@unifi.it},\\ \texttt{agil@ifae.es}, \texttt{giuliano.panico@unifi.it}}}

\thispagestyle{empty}

\newpage

\begingroup
\tableofcontents
\endgroup 

\setcounter{equation}{0}
\setcounter{footnote}{0}
\setcounter{page}{1}

\newpage

\section{Introduction}\label{sec:intro}

First order phase transitions (PhTs) in the early Universe proceed through the nucleation of bubbles of a stable phase within a metastable background.
Afterwards bubbles expand in the hot plasma and coalesce, filling the whole space.
This sequence of processes is characterised by a huge amount of energy stored in the gradients of the scalar field controlling the transition, in sound waves and turbulence in the plasma,
all of them sourcing a stochastic background of gravitational waves. 

The recent observation of gravitational waves has renewed a vivid interest in the study of the dynamics of such transitions. 
Indeed, the sensitivity regions of future experiments, such as the European interferometer LISA~\cite{Caprini:2015zlo,Caprini:2019egz}, the Japanese project DECIGO~\cite{Kawamura:2006up,Kawamura:2011zz} and the Chinese Taiji~\cite{Hu:2017mde,Ruan:2018tsw} and TianQin~\cite{TianQin:2015yph} proposals, will probe a range of the expected peak frequencies of PhTs at the electroweak (EW) scale. These interferometers will provide us with a new tool that can support collider experiments in the quest for the physics beyond the Standard Model (BSM), in particular for theories potentially affecting the dynamics of the EW symmetry breaking.

Furthermore, a stochastic gravitational wave background is not the only cosmological relic left after the completion of a PhT. 
A matter-antimatter asymmetry, dark matter remnants, primordial black holes, magnetic fields and other topological defects can also be produced.
A quantitative determination of these quantities obviously requires an accurate modelling of the PhT dynamics. 
This is controlled, among few other parameters, by the propagation velocity of the bubble wall. 

In the steady state regime, the speed of the wall is a result of the balance of the internal pressure, due to the potential difference between the two phases,
and the external friction exerted by the plasma particles impinging on the wall. In fact, the motion of a bubble drives the plasma out of equilibrium inducing a backreaction that slows down its propagation.
Despite its relevance and the huge amount of literature on the topic, see for instance~\cite{Moore:1995ua,Moore:1995si,John:2000zq,Moore:2000wx,Konstandin:2014zta,Kozaczuk:2015owa,Bodeker:2017cim,Cline:2020jre,Laurent:2020gpg,BarrosoMancha:2020fay,Hoche:2020ysm,Azatov:2020ufh,Balaji:2020yrx,Cai:2020djd,Wang:2020zlf,Friedlander:2020tnq,Cline:2021iff,Cline:2021dkf,Bigazzi:2021ucw,Ai:2021kak,Lewicki:2021pgr,Gouttenoire:2021kjv,Dorsch:2021ubz,Dorsch:2021nje,Megevand:2009gh,Espinosa:2010hh,Leitao:2010yw,Megevand:2013hwa,Huber:2013kj,Megevand:2013yua,Leitao:2014pda,Megevand:2014yua,Megevand:2014dua} for an incomplete list,
this is one of the parameters on which we have less theoretical control.

The first computation of the bubble speed in the SM can be found in the seminal work of Moore and Prokopec~\cite{Moore:1995ua,Moore:1995si} where the authors explicitly determined the friction induced by the plasma on the wall from a microphysics calculation, namely by evaluating all the relevant interactions between the plasma particles and the bubble.    
This requires the determination of the deviations from the equilibrium distributions of the different species in the plasma through the solution of the corresponding Boltzmann equations. 

Other phenomenological approaches have also been explored which rely, instead, on a parameterization of the friction in terms of a viscosity parameter~\cite{Megevand:2009gh,Espinosa:2010hh,Leitao:2010yw,Megevand:2013hwa,Huber:2013kj,Megevand:2013yua,Leitao:2014pda,Megevand:2014yua,Megevand:2014dua}.

The formalism introduced in refs.~\cite{Moore:1995ua,Moore:1995si},  that we will denote as the ``old formalism'', necessarily requires the use of an ansatz for the distribution functions. 
This is needed to parametrize their momentum dependence and, then, to compute the local collision integral of the Boltzmann equation. 
In ref.~\cite{Moore:1995si} the fluid approximation was employed assuming that the deviation from the equilibrium distributions is entirely described by only three perturbations for each species in the plasma: the chemical potential, the temperature and the velocity fluctuations. The perturbations are then extracted by taking moments of the Boltzmann equation with suitable weights. In practice, the integro-differential Boltzmann equation is converted into a much simpler system of ordinary differential equations.
By construction, the fluid approximation is equivalent to a first order expansion in the momenta of the deviation from the equilibrium distribution functions.

A peculiar feature of the fluid approximation is that the Liouville operator of the Boltzmann equation develops a zero eigenvalue at the speed of sound $c_s$ and, for larger velocities, all the perturbations trail the source term. This implies that any non-equilibrium dynamics is suppressed in front of the bubble wall~\cite{Konstandin:2014zta} with significant consequences especially for non-local EW baryogenesis which would result to be extremely inefficient for bubble walls faster than $c_s$. 

This has been the common lore for many years. But recently in refs.~\cite{Cline:2020jre,Laurent:2020gpg} it has been argued that the singularity is only an artifact of the first order truncation in momenta and of the particular set of weights chosen to extract the perturbations. Indeed, different choices of weights can shift the position of the singularity, suggesting that the speed of sound should not be a critical value for the particle diffusion as described by the fluid equations. In ref.~\cite{Laurent:2020gpg} this problem was overcome by introducing a ``new formalism'', as dubbed by the authors, which relies on a different parameterization of the non-equilibrium distributions (specifically for the velocity perturbation), different weights and a factorization ansatz~\cite{Cline:2000nw}. As a result, the new formalism wipes off the discontinuity at the speed of sound while still providing, for small velocities, quantitatively similar results to the fluid approximation. 

The same issue has also been recently revisited in ref.~\cite{Dorsch:2021ubz} for the computation of the baryon asymmetry and in ref.~\cite{Dorsch:2021nje} for the computation of the friction on the bubble wall, two problems that share many similarities. In these works the fluid approximation has been generalized by including higher orders in the small momenta expansion
and the absence of the singularity for the perturbations of the heavy species has been corroborated. Besides the issue of the singularity, large differences in the friction arise, with respect to the old formalism, when higher orders are included.
This confirms that the use of the fluid approximation, other than being not fully justified, is not particularly reliable, neither qualitatively nor quantitatively.
A major consequence is that EW-baryogenesis is indeed achievable for supersonic bubbles opening up the parameter space of many BSM models, in which the observed baryon asymmetry can be reproduced while enhancing, at the same time, the strength of the stochastic gravitational wave background.

Even though the absence of the singularity for the heavy massive species could already be inferred in ref.~\cite{Moore:1995si}, the speed of sound in the plasma still played a peculiar role in the old formalism as it provides a peak in the integrated friction for $v \simeq c_s$.\footnote{This is true for a wide range of parameters of the model, in particular if the wall thickness is not too large and the interaction strength among the plasma particles is not too strong. If these conditions are not valid, a smooth behavior can be present, as found in ref.~\cite{Dorsch:2021nje}.} The same peak (possibly accompanied by others, one for each vanishing eigenvalue of the Liouville operator) remains even if higher orders in the momenta expansion are included. As we will clarify with our analysis, such behavior is absent from the actual solution, confirming that the speed of sound is not a critical threshold of the friction for massive species.\footnote{For all the massless background species, as pointed out in ref.~\cite{Moore:1995si} and confirmed in ref.~\cite{Dorsch:2021nje} through hydrodynamic considerations~\cite{Laine:1993ey,Ignatius:1993qn,Kurki-Suonio:1995rrv,Espinosa:2010hh,Dorsch:2021nje}, a discontinuity of the temperature and fluid velocity at the bubble front for bubble velocities close to the speed of sound could be present. This discontinuity turns into a singularity of the background perturbations at $c_s$ in the linearized Boltzmann equation.}

The approaches discussed above are clearly affected by ambiguities. First of all, they all rely on an ansatz for the shape of non-equilibrium distribution functions which, both in the new and old formalisms (extended or not), is unavoidable in order to compute the collision integrals. Moreover, the choice of the basis and of the weights is not unique and different ansatzes have important qualitative and quantitative impacts on the resulting distribution functions. As such, a full solution of the Boltzmann equation that does not impose any specific momentum dependence
is necessary to provide reliable quantitative predictions for both the non-equilibrium distribution functions and the friction exerted on the bubble wall, and to clarify the issue of the presence of a singularity. 
In fact, by feeding these new results into the equation of motion of the Higgs field, one will be able to carry out a precise computation of the wall speed and of the actual profile of the domain wall (DW). This is a necessary step towards a quantitative and reliable method to asses the potential of a given BSM extension to yield interesting predictions for the relics mentioned above. This is the goal of the present work.

In this paper we will present, for the first time, a fully quantitative solution of the Boltzmann equation.
Since the absence of an ansatz prevents the direct computation of the collision integral, in order to extract the solution we will adopt an iterative method.
In particular, as it will be detailed below, the collision integral can be split into two parts, one proportional to the solution itself and another effectively treated as a source term. At each iteration, the latter is evaluated using the solution obtained at the previous step. With a clever choice of the starting solution, convergence can be reached within a very small number of steps.

For the purpose of presenting the methodology and to quantitatively asses the differences among the aforementioned formalisms, we will consider the EWPhT and we will focus on the study of top quark species, the one with the strongest coupling, among the SM particles, to the Higgs profile and, as such, the one that provides the largest contribution to the friction.
We leave for a future work the inclusion of the electroweak gauge bosons and of the background species. 

The paper is organized as follows: in section~\ref{sec:Boltzmann} we describe our method while in section~\ref{sec:numerical} we present and discuss the numerical results.
In section~\ref{sec:conclusions} we give our conclusions and discuss future directions. All the technicalities related to the computation of the collision integrals are discussed in  Appendix~\ref{app:analytic}.

\section{The Boltzmann equation}\label{sec:Boltzmann}

As discussed in the Introduction, our goal is to determine the solution of the Botzmann equation for the distribution function of
the plasma in the presence of an expanding bubble of true vacuum. Once the bubble reaches a radius much larger than the thickness of
its wall, to a good approximation we can adopt the planar limit, considering a flat DW with a velocity parallel to its normal vector.

Assuming that (for long enough time) a steady state is reached, it is convenient to write the Boltzmann equation in the wall frame
(i.e. the frame in which the DW is at rest) in which the solutions are stationary. Orienting the $z$ axis along the velocity of the
DW, the equation for the distribution function $f$ of a particle species in the plasma is 
\begin{equation}
{\cal L}[f] \equiv \left(\frac{p_z}{E} \partial_z - \frac{(m^2(z))'}{2E} \partial_{p_z}\right) f = - {\cal C}[f]\,,
\end{equation}
where $m(z)$ is the mass of the particle, which in general depends on the position $z$, and $(m^2)' \equiv d m^2/dz$.
The term ${\cal C}$ appearing in the right hand side of the equation is the collision integral, describing local microscopic
interactions among the plasma particles, while ${\cal L}$ is the Liouville operator.

The collision term ensures that far from the DW, where the forces acting on the system basically vanish, each particle species approaches
local thermal equilibrium. In the presence of a background fluid with a large number of degrees of freedom (in our case given
by gluons and light quarks, which are not much affected by the Higgs phase transition), we can assume that the local thermal
equilibrium is described by the standard Fermi or Bose--Einstein distributions for a fluid moving with velocity $v$ along the
$z$ axis,\footnote{This corresponds to choose the DW to move with velocity $-v$ along the $z$ direction in the plasma frame.}
namely
\begin{equation}
f_v = \frac{1}{e^{\beta \gamma(E - v p_z)} \pm 1}\,,
\end{equation}
with $\beta = 1/T$ and $\gamma = 1/\sqrt{1- v^2}$.

Deviations with respect to the local equilibrium distribution are present mostly close to the DW and are expected to vanish for
$z \to \pm \infty$. For small perturbations, the distribution function can be written as
$f = f_v + \delta f$ and the Boltzmann equation can be linearized in $\delta f$:
\begin{equation}\label{eq:Boltz_lin}
\left(\frac{p_z}{E} \partial_z - \frac{(m(z)^2)'}{2E} \partial_{p_z}\right) \delta f + {\overline{\cal C}}[\delta f] = \frac{(m(z)^2)'}{2E} \partial_{p_z} f_v = \beta \gamma v \frac{(m(z)^2)'}{2E} f_v'\,,
\end{equation}
where we defined
\begin{equation}
f_v' \equiv - \frac{e^{\beta \gamma(E- v p_z)}}{(e^{\beta \gamma(E- v p_z)} \pm 1)^2}\,,
\end{equation}
and ${\overline{\cal C}}[\delta f]$ denotes the collision integral linearized in $\delta f$.
Notice that the only source term in the linearized Boltzmann equation comes from the Liouville operator ${\cal L}$ applied to the
local equilibrium distribution. The collision integral, on the contrary, vanishes when computed on $f_v$, ${\cal C}[f_v] = 0$.
Sizable values for the source term are therefore present only close to the DW, where the non-trivial Higgs profile
generates a non-negligible $z$ dependence in $m(z)$. Away from the DW, the Higgs profile is instead almost constant, thus giving
$(m^2)' \simeq 0$ and suppressing the source term. This behavior is in agreement with the naive expectation that deviations from
local thermal equilibrium are only present close to the DW and should decrease to zero away from it.

\subsection{Flow paths and the Liouville operator}\label{sec:Liouville}

As a first step towards finding a solution of the Boltzmann equation, we need to rewrite the Liouville differential operator in a simpler
form.\footnote{The strategy we use to rewrite the Boltzmann equation is the well-known ``method of characteristics''
for first-order partial differential equations.}
It is straightforward to check that, along the paths on which both the transverse momentum\footnote{That is the component of the momentum parallel to the DW.} $p_\bot$ and the quantity $p_z^2 + m^2(z)$ are constant, the differential operator simply reduces to a total derivative with respect to $z$:
\begin{equation}
{\cal L} = \left(\frac{p_z}{E} \partial_z - \frac{(m^2(z))'}{2E} \partial_{p_z}\right) \quad \to \quad \frac{p_z}{E} \frac{d}{dz}\,.
\end{equation}

The physical interpretation of the paths is quite intuitive. They correspond to the trajectories of the particles in the $(p_\bot, p_z, z)$
phase space in the collisionless limit. In this limit the energy of the particles and their momentum parallel to the DW are conserved
(due to time invariance and translation invariance along the DW), therefore the trajectories of the particles are given by
\begin{equation}
\left\{\hspace{-.25em}
\begin{array}{l}
E = \sqrt{p_\bot^2 + p_z^2 + m^2(z)} = const\\
\rule{0pt}{1.5em}p_\bot = const
\end{array}
\right.
\quad\Rightarrow\quad
p_z^2 + m(z)^2 = const\,.
\end{equation}

\begin{figure}
\centering
\includegraphics[width=.52\textwidth]{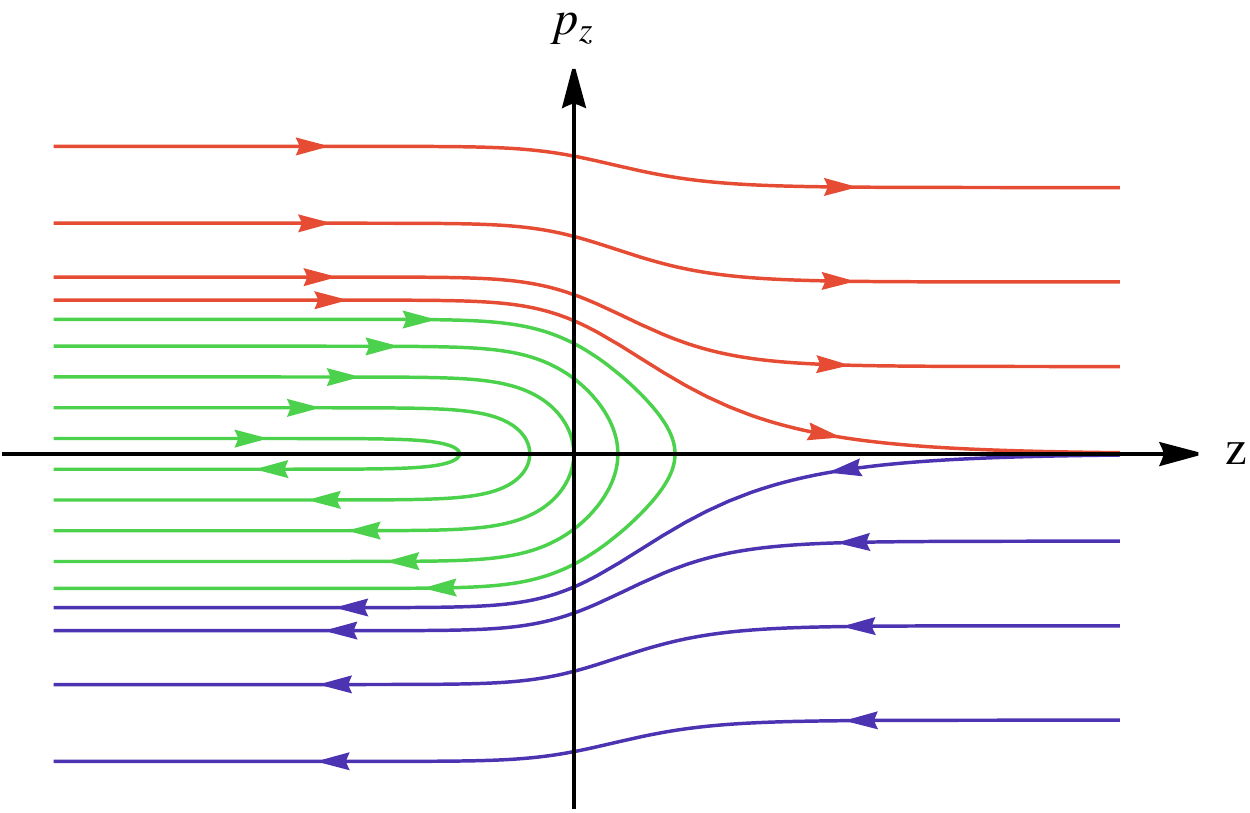}
\hfill
\raisebox{1em}{\includegraphics[width=.43\textwidth]{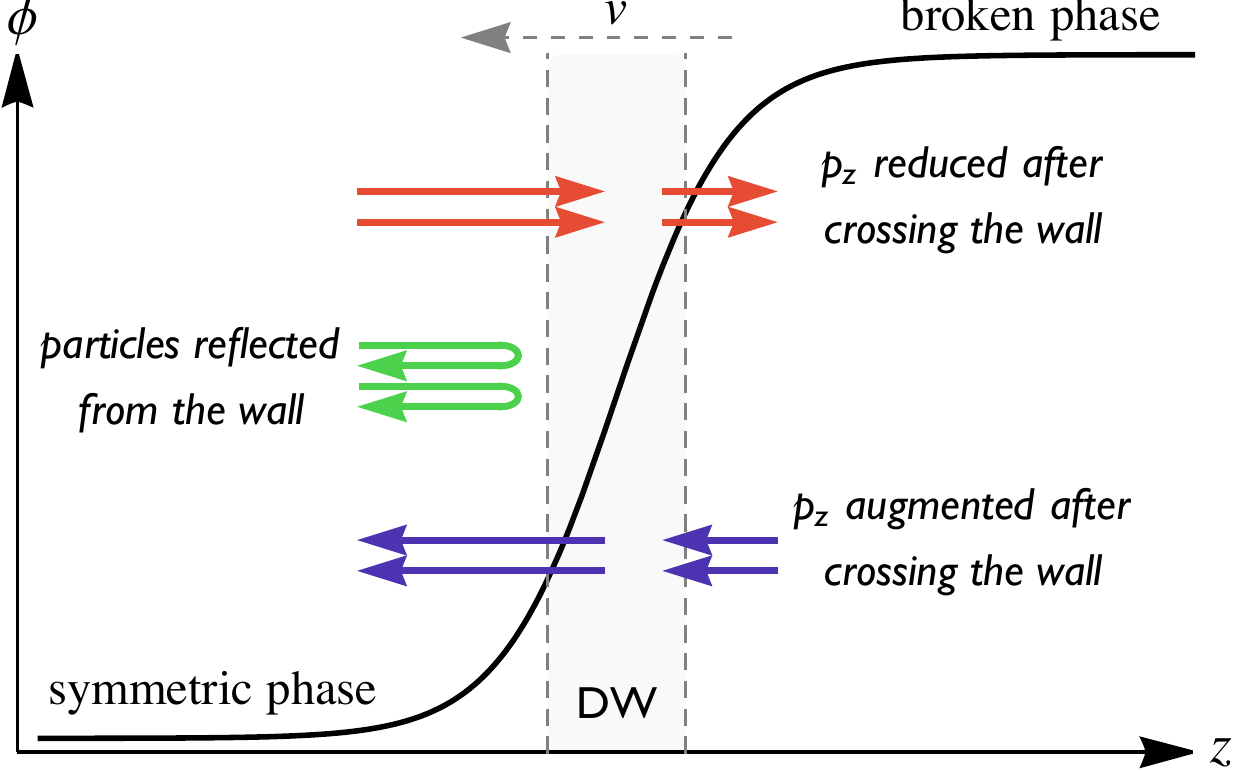}}
\caption{Left panel: Paths with fixed energy and transverse momentum in the $z - p_z$ phase space for the choice $m(z) \propto 1 + \tanh(z/L)$. The red, green and purple colors denote sets of contours with different behavior. The arrows show the flow of
a particle within the phase space. Right panel: Schematic representation of the behavior of the particles across the DW.}\label{fig:flow_paths}
\end{figure}

The condition $p_z^2 + m^2(z) = const$ gives rise to different classes of \emph{flow paths}. Since the mass of the particle species
receives a contribution from the Higgs VEV, we expect it to smoothly increase going from the symmetric phase outside the bubble
to the symmetry-broken one inside it. In particular, if we are interested in particles whose mass comes entirely from EW symmetry
breaking (as it happens for the top quark and for the $W$ and $Z$ bosons), we can assume that $m(z) \to 0$ for $z \to -\infty$,
while it approaches a constant value $m(z) \to m_0 > 0$ for $z \to +\infty$.
In this case three types of flow paths are present:
\begin{itemize}
\item[i)] for $p_z(-\infty) \geq m_0$ the path goes from $z = -\infty$ to $z = +\infty$ and has always $p_z > 0$,
\item[ii)] for $-m_0 < p_z(-\infty) < m_0$ the path goes from $z = -\infty$ to the point $\bar z$ in which $p_z(\bar z) = 0$ (i.e.~the point that solves the equation $m(\bar z) = p_z(-\infty)$) and then goes back to $z = -\infty$,
\item[iii)] for $p_z(-\infty) \leq -m_0$ the path goes from $z = +\infty$ to $z = -\infty$ and has always $p_z < 0$.
\end{itemize}
The three classes of curves are shown schematically in fig.~\ref{fig:flow_paths} for the choice $m(z) \propto 1 + \tanh(z/L)$, with $L$ denoting the wall thickness.
The paths of type i, ii and iii correspond to the red, green and purple curves respectively.\footnote{We stress that the approximation, typically used in the literature, in which the $(m^2)'/(2E) \partial_{p_z} \delta f$ term
is neglected in the Boltzmann equation could lead to an inaccurate result in our approach. Neglecting that term, in fact, modifies
the flow paths making all of them straight lines with fixed $p_z$. This completely changes the shape of the curves
in the region $|p_z| \leq m_0$, thus potentially giving a very different solution of the equation.
}

Exploiting the flow paths we can straightforwardly solve any differential equation of the form
\begin{equation}\label{eq:boltz_S}
{\cal L}[\delta f] - \frac{1}{E} {\cal Q}\, \delta f = \frac{p_z}{E}{\cal S}\,,
\end{equation}
where ${\cal Q}$ and ${\cal S}$ are generic functions of $E$, $p_z$ and $z$,
and the factors $1/E$ and $p_z/E$ have been chosen for convenience. Rewriting
the above equation along the flow paths we find
\begin{equation}
\left(\frac{d}{dz} - \frac{\cal Q}{p_z}\right) \delta f = {\cal S}\,,
\end{equation}
whose general solution is
\begin{equation}\label{eq:res_S}
\delta f = \left[B(p_\bot, p_z^2 + m(z)^2) + \int_{\bar z}^z e^{-{\cal W}(z')} {\cal S}\, dz'\right]e^{{\cal W}(z)}\,,
\end{equation}
where ${\cal W}$ is given by
\begin{equation}
{\cal W}(z) = \int^z \frac{\cal Q}{p_z} dz'\,,
\end{equation}
and all the integrals are evaluated along the flow paths. Notice that the lower integration boundary in the definition of
${\cal W}$ can be freely chosen (for each flow path) without affecting the result in eq.~(\ref{eq:res_S}).

The function $B(p_\bot, p_z^2 + m(z)^2)$, which is constant along the flow paths, is arbitrary and can be fixed by enforcing
the required boundary conditions. Let us focus separately on the three classes of flow paths.
\begin{itemize}
\item[i)] The first type of paths describes particles that travel in the positive $z$ direction, and eventually enter into the bubble. It is natural
to choose the boundary conditions in such way that $\delta f$ vanishes at $z \to -\infty$, that is well before the particle hits the DW.
This can be enforced by choosing
\begin{equation}
\delta f = \left[\int_{-\infty}^z e^{-{\cal W}}\,{\cal S}\, dz'\right]e^{{\cal W}(z)}\,.
\end{equation}

\item[ii)] The second type of paths describes particles that initially travel in the positive $z$ direction, hit the DW and are reflected.
It is natural to choose the boundary conditions similarly to what we did for the previous type of paths. Therefore we have
\begin{equation}
\delta f = \left[\int_{-\infty_\uparrow}^z e^{-{\cal W}}\,{\cal S}\, dz'\right]e^{{\cal W}(z)}\,,
\end{equation}
where the up arrow in the lower integration boundary indicates that the integration is performed starting from $z \to -\infty$
in the half path with $p_z > 0$.

\item[iii)] The third type of paths describes particles that travel in the negative $z$ direction, and eventually exit from the bubble. We can choose the boundary conditions in such way that $\delta f$ vanishes at $z \to +\infty$, that is well before the particles exit from the bubble. This can be obtained by choosing
\begin{equation}
\delta f = - \left[\int_z^{+\infty} e^{-{\cal W}}\,{\cal S}\, dz'\right]e^{{\cal W}(z)}\,.
\end{equation}
\end{itemize}
The consistency of all these solutions requires ${\cal Q} < 0$. We verified numerically that this condition is satisfied for the
equations we are considering.

The form of the solution clearly shows the role of the term $({\cal Q}/E) \delta f$ in driving the system towards the local thermal
equilibrium, i.e.~in decreasing the value of $\delta f$. In fact, due to the exponential factors, the impact of the source term ${\cal S}$
is exponentially suppressed with the distance. The decay length is of order $\ell \sim p_z/{\cal Q}$ and, as expected,
decreases for larger values of the collision term.

\subsection{Finding a solution for the Boltzmann equation}\label{sec:solution}

Although the full Boltzmann equation is not of the form of eq.~(\ref{eq:boltz_S}), we can use the latter to implement an approximation
by steps. The basic idea is to split the collision integral $\overline{\cal C}[\delta f]$ in two pieces: a term analogous to $({\cal Q}/E) \delta f$ in eq.~(\ref{eq:boltz_S}), and a second term that is included in the source term ${\cal S}$ and is used to correct the solution through iterations.

Let us now analyze in details the collision integral. For simplicity we consider the collision term for the $2 \to 2$ processes
of a single particle species, but the general case can be treated in an analogous way.
The collision integral is given by
\begin{equation}
{\cal C}[f] = \sum_i \frac{1}{4N_p E_p} \int\! \frac{d^3{\bf k}\,d^3{\bf p'}\,d^3{\bf k'}}{(2\pi)^5 2 E_k 2E_{p'} 2E_{k'}} |{\cal M}_i|^2
\delta^4(p+k-p'-k') {\cal P}[f]\,,
\end{equation}
with
\begin{equation}
{\cal P}[f] = f(p) f(k)(1 \pm f(p'))(1 \pm f(k')) - f(p') f(k')(1 \pm f(p))(1 \pm f(k))\,,
\end{equation}
where the sum is performed over all the relevant scattering processes, whose squared scattering amplitude is $|{\cal M}_i|^2$.
In the above formula $N_p$ is the number of degrees of freedom of the incoming particle with momentum $p$, $k$ is the momentum of the second incoming particle, while $p'$ and $k'$ are the momenta of the outgoing particles.
The $\pm$ signs are $+$ for bosons and $-$ for fermions.

From the above expression we can easily derive the collision integral for the linearized Boltzmann equation. As a consequence of the
conservation of the total $4$-momentum in the collision processes we have that for the local equilibrium distribution ${\cal C}[f_v] = 0$.
Moreover the linear terms in $\delta f$ can be expressed as
\begin{equation}\label{eq:P_bar}
\overline{\cal P} = f_v(p) f_v(k) (1 \pm f_v(p')) (1 \pm f_v(k')) \sum \frac{\mp \delta f}{f_v'}\,,
\end{equation}
where the $-(+)$ sign in the sum applies to incoming (outgoing) particles.

The $\overline{\cal C}[\delta f]$ collision integral can therefore be split in two parts. One of them depends only on $\delta f(p)$
and is given by
\begin{equation}
\frac{-f_v(p)}{f_v'(p)} \delta f(p) \sum_i \frac{1}{4N_p E_p} \int\! \frac{d^3{\bf k}\,d^3{\bf p'}\,d^3{\bf k'}}{(2\pi)^5 2 E_k 2E_{p'} 2E_{k'}} |{\cal M}_i|^2
\delta^4(p+k-p'-k') f_v(k) (1 \pm f_v(p')) (1 \pm f_v(k'))\,.
\end{equation}
This expression is clearly analogous to the term $({\cal Q}/{E}) \delta f$ in eq.~(\ref{eq:boltz_S}). The second part of the collision integral includes the terms in which $\delta f$ depends on $k$, $p'$ or $k'$ and thus appears under the integral sign.
We collectively denote these terms by $\langle \delta f \rangle$. The numerical determination of the various contributions to the collision integral can be drastically simplified through
a clever choice of integration variables. The explicit procedure is explained in Appendix~\ref{app:analytic}.

In order to numerically solve the Boltzmann equation, a possible strategy is to formally rewrite it as eq.~(\ref{eq:boltz_S}) by
including $\langle \delta f\rangle$ in the source term ${\cal S}$. The solution can then be found by iteration, inserting into the
equation the value of $\langle \delta f\rangle$ obtained by using the solution at the previous step.

\section{Numerical analysis}\label{sec:numerical}

In this section we apply the iterative method explained above to numerically solve the Boltzmann equation.
For simplicity we focus on a single species in the plasma, the top quark, which is the state with largest coupling to the Higgs
and is thus expected to provide one of the most relevant effects controlling the DW dynamics. The analysis of the top quark
distribution should be sufficient to provide a robust assessment of the plasma dynamics and to obtain an indication of how much
the weighted method used in the literature to solve the Boltzmann equation is qualitatively and quantitatively accurate.
We leave for future work the inclusion of the contributions from the $W$ and $Z$ bosons, which are expected to be roughly
of the same size as the top quark ones.

The iterative approach explained in the previous section could be straightforwardly applied to determine the solution of the Boltzmann
equation. However, in order to improve the convergence of the iterative steps, a slight modified procedure proves
more convenient. Since the separation of the collision integral into a contribution to $({\cal Q}/{E}) \delta f$ and a contribution
to ${\cal S}$ is to a large extent arbitrary, we can devise a splitting that helps in reducing as much as possible the source term.

\subsection{Annihilation only}

Focusing on the top quark case, it can be shown that the main contribution to the collision integral comes from the annihilation
process $t \bar t \to g g$, whereas the scattering of tops on gluons and light quarks gives smaller contributions. In our numerical analysis we will therefore consider at first only the contribution from annihilation, including scattering effects afterwards.

In the annihilation case, the linear terms in the perturbation $\delta f$ appear in the following combination (see eq.~(\ref{eq:P_bar}))
\begin{equation}
\overline{\cal P} = f_v(p) f_v(k) (1 + f^g_v(p')) (1 + f^g_v(k')) \left(- \frac{\delta f(p)}{f_v'(p)} - \frac{\delta f(k)}{f_v'(k)}\right)\,,
\end{equation}
where $f^g_v$ denotes the equilibrium distribution for the background gluons (which is approximately unperturbed
since the number of degrees of freedom in the background species is large). In the above formula the $\delta f(p)$ and $\delta f(k)$
terms play an analogous role, but their effects become distinct in the collision integral since an integration over $k$ is performed.
It is nevertheless evident that the impact of the $\delta f(k)$ term in the Boltzmann equation is not particularly suppressed, as can be
understood averaging the equation by integration over $p$, in which case the $\delta f(p)$ and $\delta f(k)$ become
exactly equal. This line of reasoning suggests that treating the $\delta f(k)$ contribution as source term,
while including the $\delta f(p)$ term in $({\cal Q}/p_z) \delta f$ might lead to slow convergence. To overcome this difficulty we will
use a slightly modified procedure. We rewrite $\overline{\cal P}$ as
\begin{equation}
\overline{\cal P} = f_v(p) f_v(k) (1 + f^g_v(p')) (1 + f^g_v(k')) \left[-2 \frac{\delta f(p)}{f_v'(p)} + \left(\frac{\delta f(p)}{f_v'(p)} - \frac{\delta f(k)}{f_v'(k)}\right)\right]\,,
\end{equation}
and we interpret the first contribution as $({\cal Q}/p_z) \delta f$, while the second one (the one in round parentheses) is treated as
a source. In this way the contribution to the source is partially canceled and faster convergence is achieved.\footnote{We also checked that, in the weighted approach to the solution of the Boltzmann equation, doubling the contribution of $\delta f(p)$ and neglecting $\delta f(k)$ gives a fair approximation of the exact result.}

To determine the numerical solution of the Boltzmann equation we used a dedicated C++ code, validating the results with Mathematica~\cite{Mathematica}.
The solution was computed on a three-dimensional grid in the variables $z$, $p_\bot$ and $p_z$ restricted to the
intervals $z/L \in [-7,7]$,\footnote{The vanishing boundary conditions on the solution were imposed at the boundaries
of the considered region. We verified that this choice does not introduce a significant distortion of the solution.}
$p_\bot/T \in [0, 15]$, and $p_z/T \in [-15, +15]$. The solution was computed on a grid with $50 \times 300 \times 100$ points, which was further refined in the region $p_\bot/T < 1$ and $|p_z|\sim m_0$, where the solution showed a fast-varying behavior. Convergence of the solution (at the $\sim 0.1\%$ level) was achieved within three iterative steps for all values of the wall velocity.
We modeled the bubble wall assuming that the Higgs profile has the following functional dependence on $z$~\cite{PhysRevD.101.063525}:
\begin{equation}
\phi(z) = \frac{\phi_0}{2}[1 + \tanh(z/L)]\,,
\end{equation}
where $L = 5/T$ is the thickness of the bubble wall and  $\phi_0 = 150\;{\rm GeV}$ is the Higgs VEV in the broken phase. We fixed the
phase transition temperature to $T = 100\;{\rm GeV}$. This choice of parameters, as we will see, determines the presence of friction peaks in the old formalism solution. It is thus well suited for differentiating the various formalisms and highlights the differences among them.

\begin{figure}
	\centering
	{\includegraphics[width=.47\textwidth]{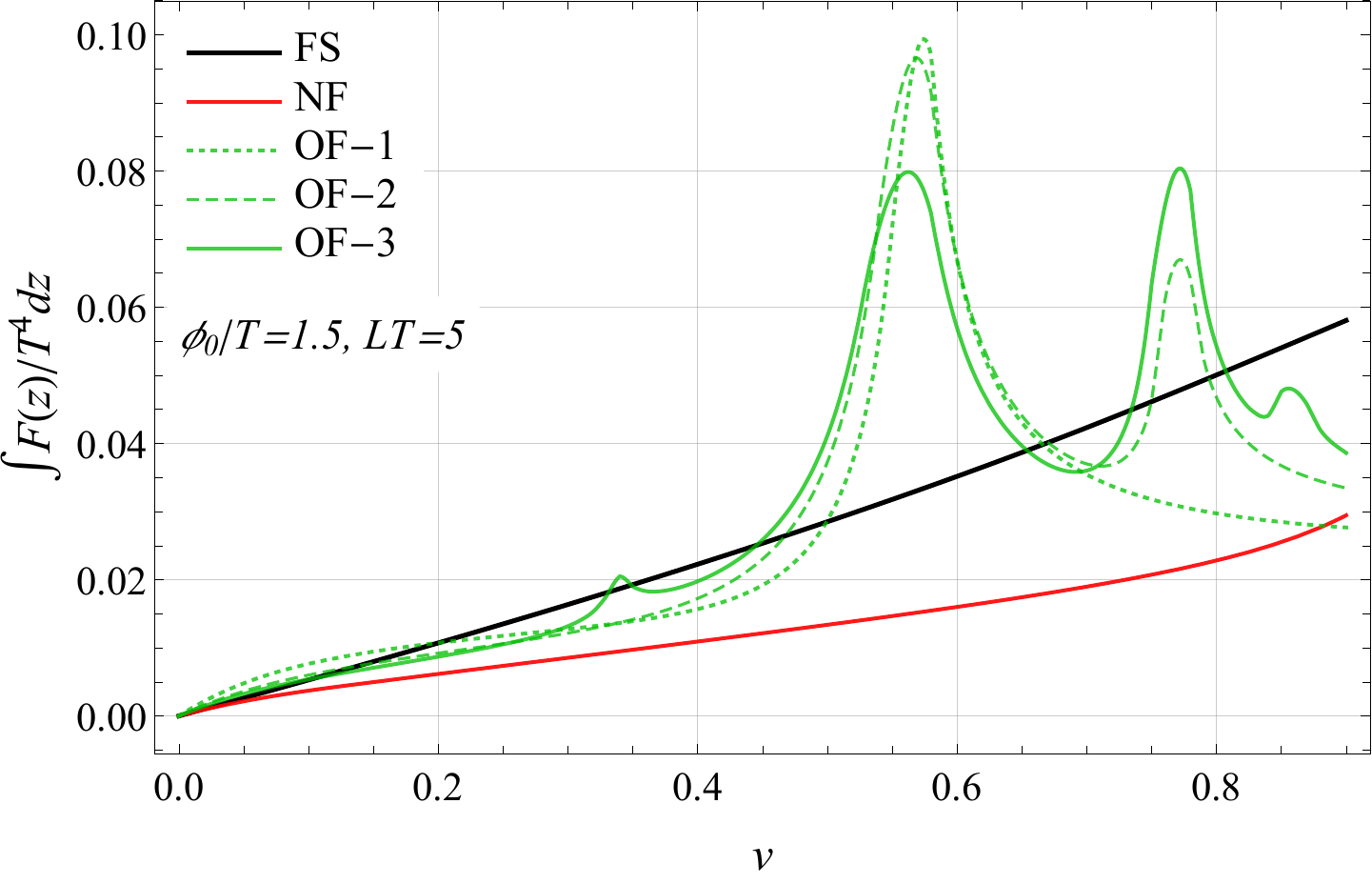}}
	\hfill
	{\includegraphics[width=.47\textwidth]{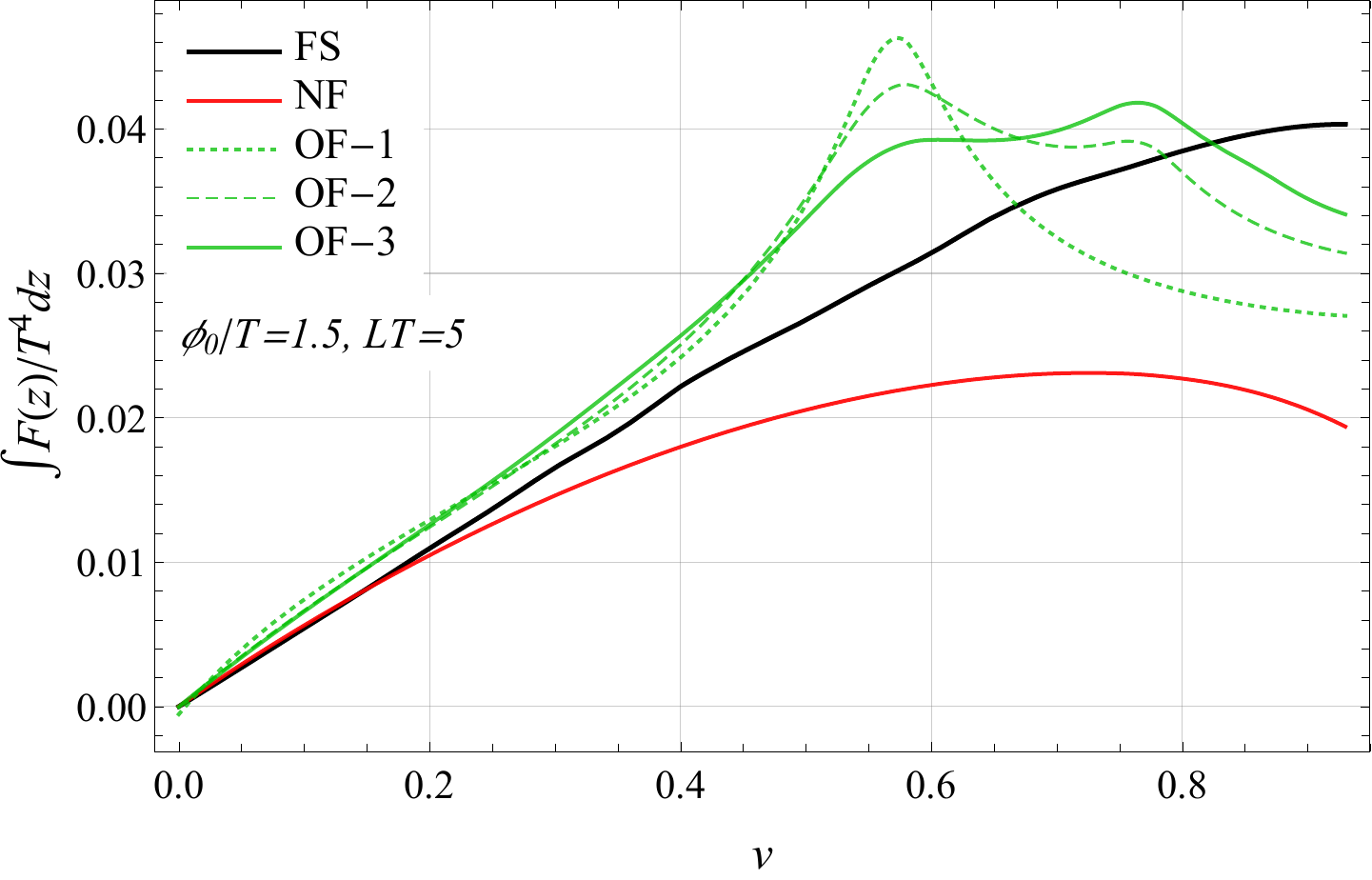}}
	\caption{Friction acting on the bubble wall as a function of the velocity. In the left plot only the top annihilation channel has been taken into account, while in the right one both annihilation and scattering are considered. The black solid line corresponds to the solution of the full Boltzmann equation (FS, our result), the dotted, dashed and solid green lines are obtained with the old formalism (OF) at order $1$, $2$ and $3$ respectively~\cite{Moore:1995si,Dorsch:2021ubz}, while the red line corresponds to the new formalism (NF)~\cite{Laurent:2020gpg}.}\label{fig:friction}
\end{figure}

An important quantity that can be derived from the numerical solution is the friction acting on the domain wall, which corresponds to the
expression~\cite{Moore:1995si}
\begin{equation}
F(z) = \frac{d m^2}{d z} N \int \frac{d^3 {\bf p}}{(2 \pi)^3 2E} \delta f(p)\,,
\end{equation}
where $N$ denotes the number of degrees of freedom ($N = 12$ for the top/antitop quark system).
In the left panel of fig.~\ref{fig:friction} we show the friction integrated over $z$ as a function of the wall velocity (solid black line).
The total friction shows a smooth behavior with a (nearly) linear growth as a function of the wall velocity.

In the same plot we compare our result with the ones obtained with the weighted methods.
In particular the green lines correspond to the total friction computed in the old formalism (OF) of ref.~\cite{Moore:1995si},
taking also into account higher-order terms in the fluid approximation~\cite{Dorsch:2021ubz}. The old formalism results
at order $1$, $2$ and $3$ are given by the dotted, dashed and solid lines respectively. The solid red line, instead, is obtained using the
new formalism (NF) of ref.~\cite{Laurent:2020gpg}.

Our result for small and intermediate velocities, $v \lesssim 0.5$ is in fair numerical agreement with the old formalism ones, which
show a minor dependence on the order used for the computation. At higher velocities, instead, the old formalism develops some peaks
related to the speed of sound in the plasma and to any other zero eigenvalue of the Liouville operator. The number of peaks and their shape crucially depend on the approximation order,
denoting an intrinsic instability of the old formalism method.\footnote{Notice that the total friction shows a continuous behavior across
the sound speed thresholds, whereas in ref.~\cite{Moore:1995si} a divergence was found.
This difference was expected, since the discontinuity found in ref.~\cite{Moore:1995si} is induced by the background contributions, which are not included in our analysis.} Our results for the full solution of the Boltzmann equation show that the peaks are an artifact of
the old formalism approach and that no strong effect is present in the top contributions for velocities close to the sound speed one.

On the other hand, the new formalism correctly predicts a smooth behavior for the total friction for all domain wall velocities.
A roughly linear dependence on $v$ is obtained up to $v \simeq 0.8$, while for larger values a faster growth is found,
in contrast with the behavior of the full solution (FS) result.
The quantitative agreement with the full solution is good only for very low velocities, $v \lesssim 0.1$, while order $50\%$ differences
can be seen for higher velocities.

\begin{figure}
	\centering
	{\includegraphics[width=.47\textwidth]{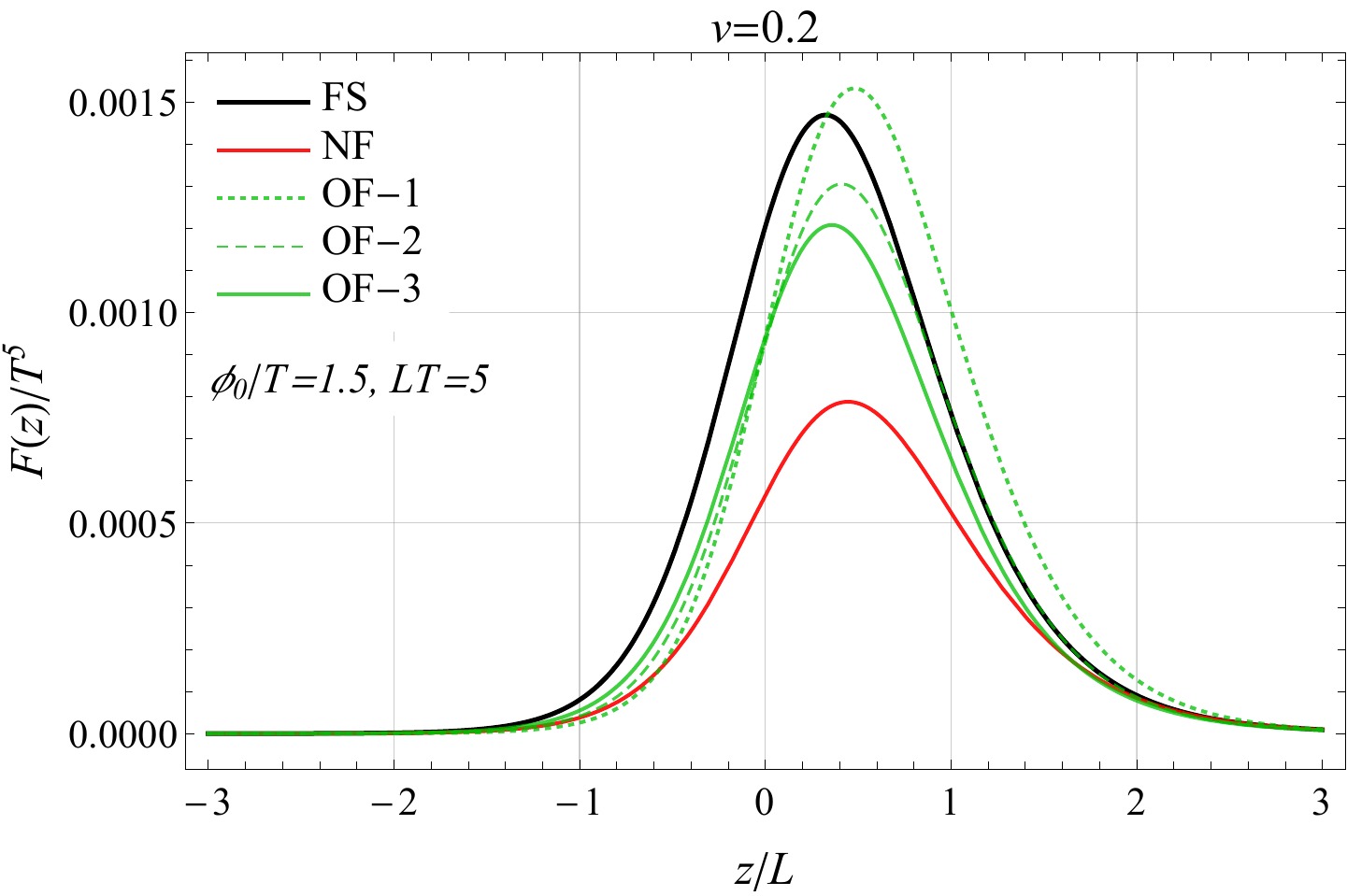}}
	\hfill
	{\includegraphics[width=.47\textwidth]{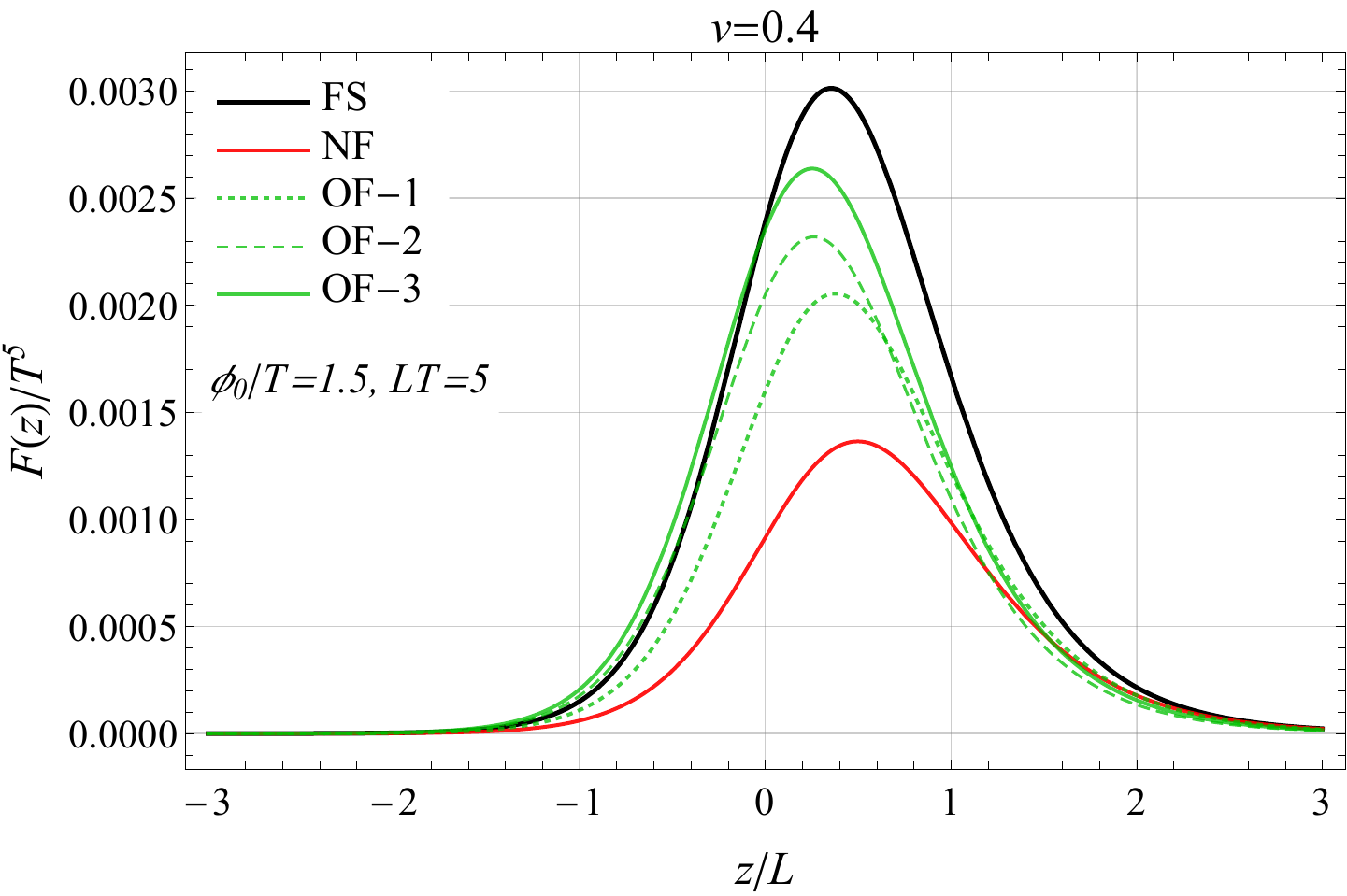}}\\
	\vspace{.5em}
	{\includegraphics[width=.47\textwidth]{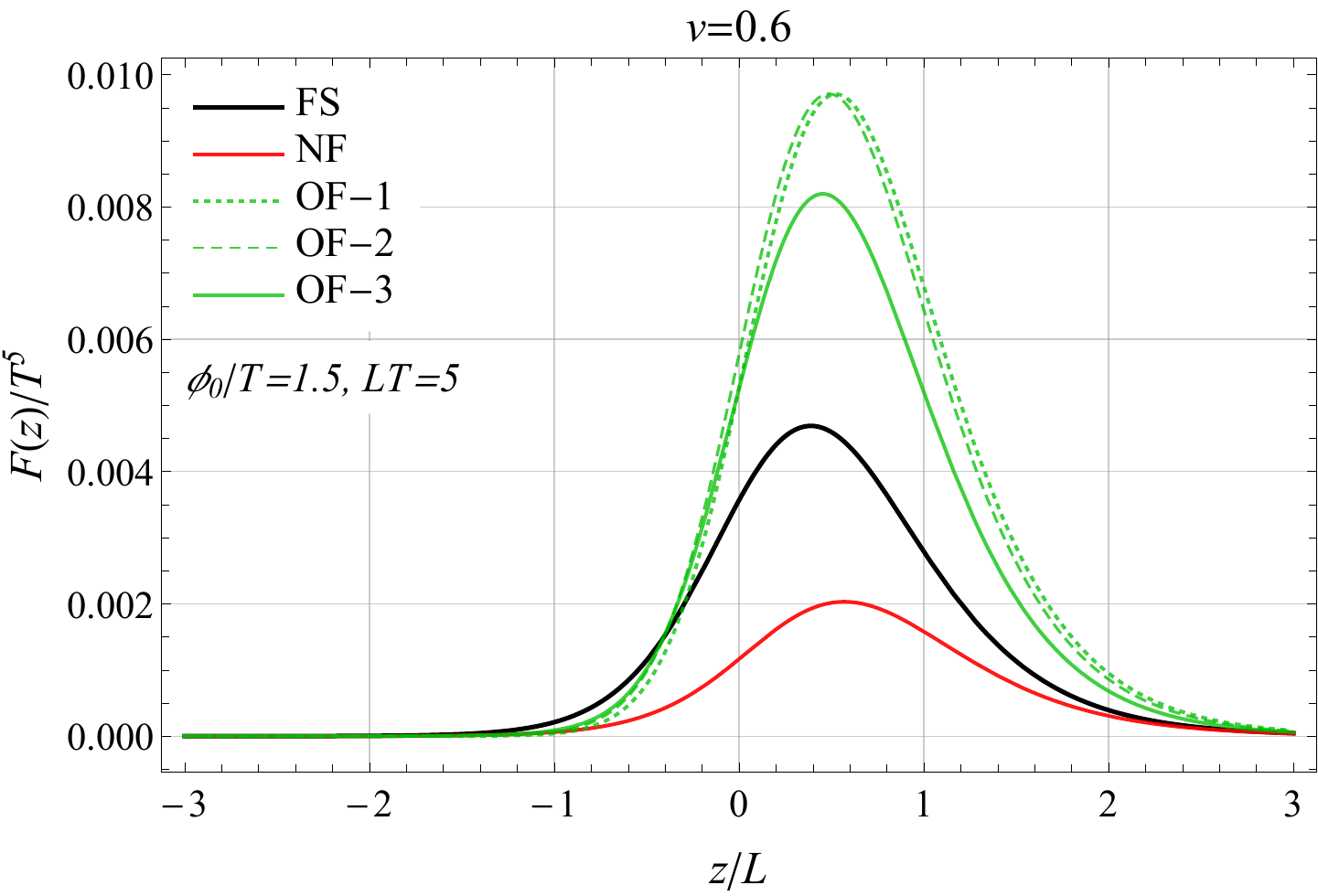}}
	\hfill
	{\includegraphics[width=.47\textwidth]{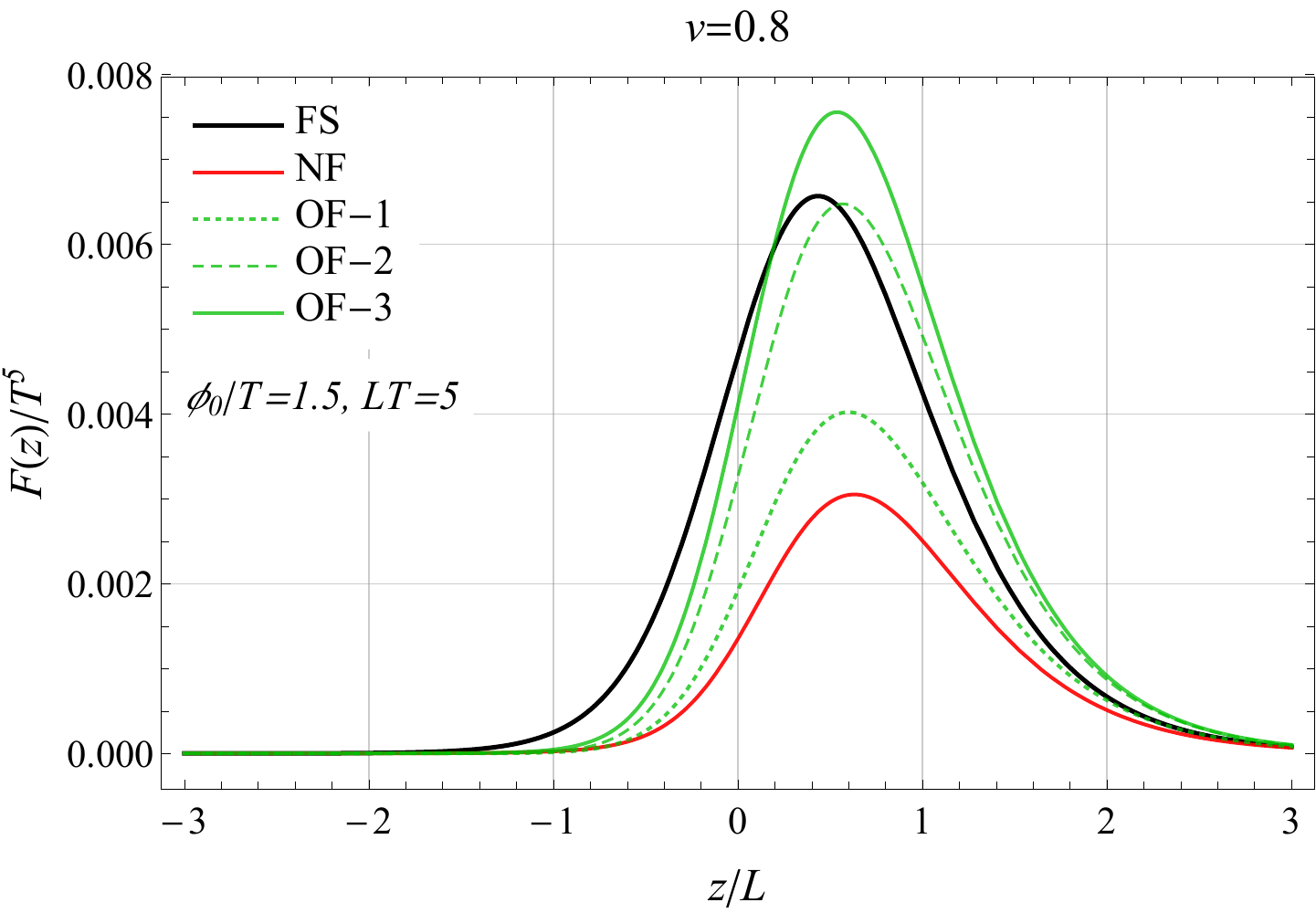}}
	\caption{Friction as a function of the position $z$ when only top annihilation processes are taken into account. The plots correspond to the wall velocities $v  = 0.2, 0.4, 0.6, 0.8$.}\label{fig:friction_ann}
\end{figure}

For a more refined comparison of the results we show in fig.~\ref{fig:friction_ann} the behavior of the friction $F(z)$ as a function
of the position. The plots clearly show that the overall shape of the friction is very similar in all approaches, the main difference being
the height of the peak. This property is not unexpected, since the size of the perturbation $\delta f$ is controlled by the
source term in the Boltzmann equation, whose $z$ dependence is given by $d m^2/dz$. One can easily check that the shape of
all the curves in the plots roughly agree with the function $d \phi^2(z)/d z$.

A more detailed comparison of $\delta f$ as a function of $z$, $p_\bot$ and $p_z$ shows drastic differences among all the approaches.
Although the overall size of $\delta f$ is comparable in all formalisms (being controlled by the source term), the various solutions
significantly differ even at the qualitative level in most of the kinematic regions. We conclude from this comparison that the fluid
approximation is not reliable if we include in the Boltzmann equation only the top annihilation channel. We will see in the following that,
introducing the top scattering processes, a better agreement is found.

\subsection{Full solution}

We now consider the Boltzmann equation for the top quark distribution, including in the collision term also the main top scattering processes, namely the ones onto gluons $t g \to t g$ and onto light quarks $t q \to t q$. We found convenient to include these additional contributions treating them as source terms in the iterative steps.

To determine the numerical solution we used a grid analogous to the one described in the annihilation-only case. The convergence of the iterative procedure is somewhat slower when top scattering processes are taken into account. For $v \leq 0.6$ we used the
solution of the annihilation-only case as starting ansatz and we performed six iterative steps to reach a good convergence.
For higher velocities the annihilation-only solution is not a convenient choice for the first iterative step, thus we started from the
full solution determined for a lower value of $v$. Also in this case six iterations were sufficient to achieve convergence.

We found that the scattering processes significantly modify the solution of the Boltzmann equation, especially for large values
of the domain wall velocity ($v \gtrsim 0.5$). The impact on the total friction acting on the domain wall is shown in the right panel
of fig.~\ref{fig:friction}. Analogously to the annihilation-only case, an almost linear dependence on the wall velocity is present
for small and intermediate $v$ values, but a flattening is present at higher velocities. Quantitatively,
the scattering processes induce only minor corrections to the total friction for $v \lesssim 0.6$, while a decrease of order $25\%$
is found for $v \sim 0.8$.

The impact of the scattering processes on the solution obtained through the weighted methods is, on the contrary, much more
pronounced. The old formalism approach (green lines in fig.~\ref{fig:friction}) including only the lowest-order perturbations predicts
a strong peak for $v \simeq 0.55$. The peak however gets substantially reduced once higher-order perturbations are included in the
expansion, with a milder additional peak forming for $v \simeq 0.75$. We expect that including additional higher-order perturbations
could smoothen the curve, giving a qualitative behavior similar to the one we get with the full solution, with a linear behavior for $v \lesssim 0.6$. At the quantitative level, however, the old formalism solution differs from the one we found by
order $10 - 25\%$.

The result obtained through the new formalism (red line in fig.~\ref{fig:friction}) is also substantially modified by the scattering contributions. In particular the increase in the friction for $v \gtrsim 0.8$ is removed and a maximum followed by a mild decrease is now found
for $v \gtrsim 0.6$. The new formalism prediction is now in good quantitative agreement with our result for $v \lesssim 0.2$,
while differences  up to order $50\%$ are found for larger velocities.

\begin{figure}
	\centering
	{\includegraphics[width=.47\textwidth]{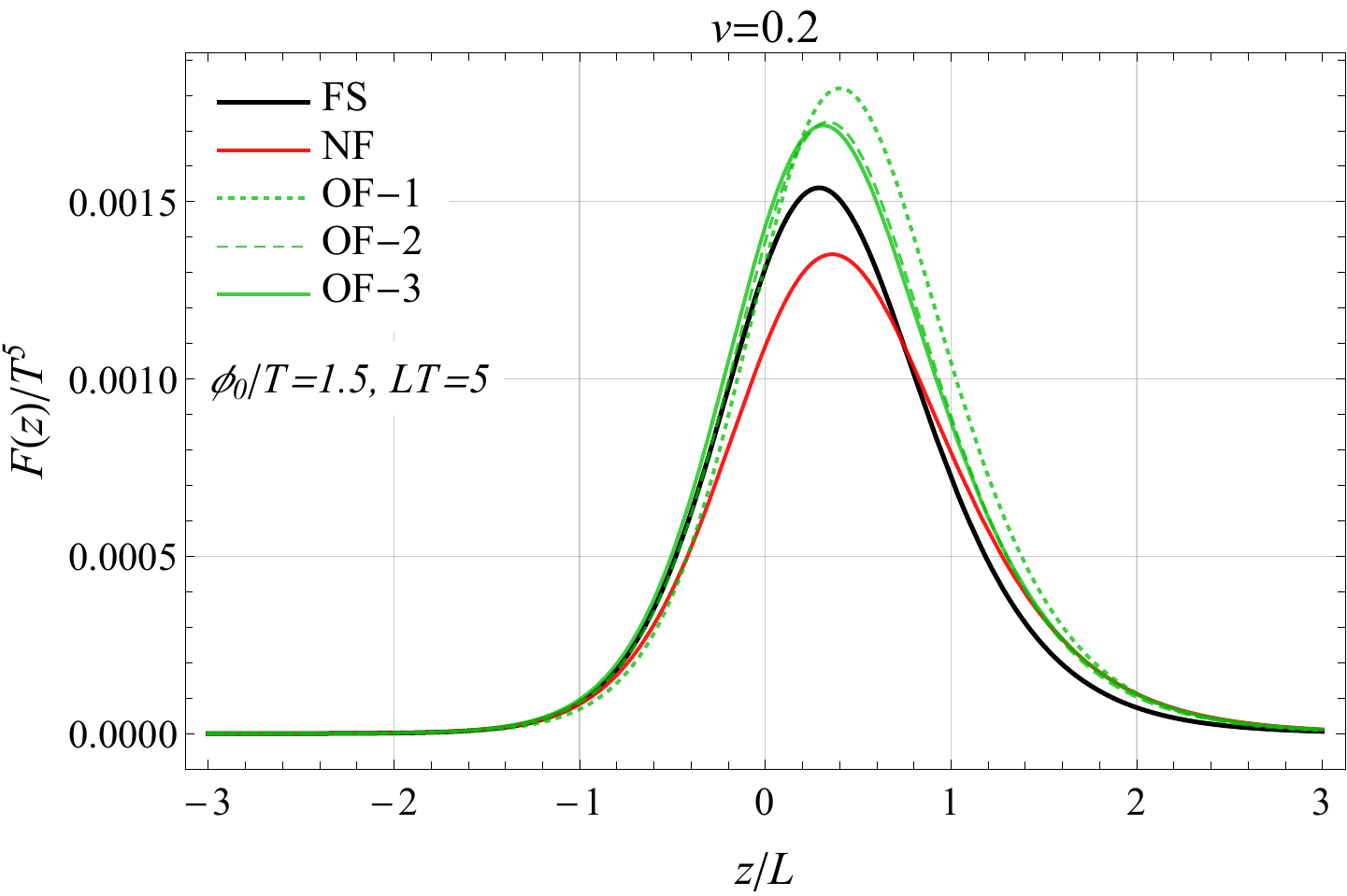}}
	\hfill
	{\includegraphics[width=.47\textwidth]{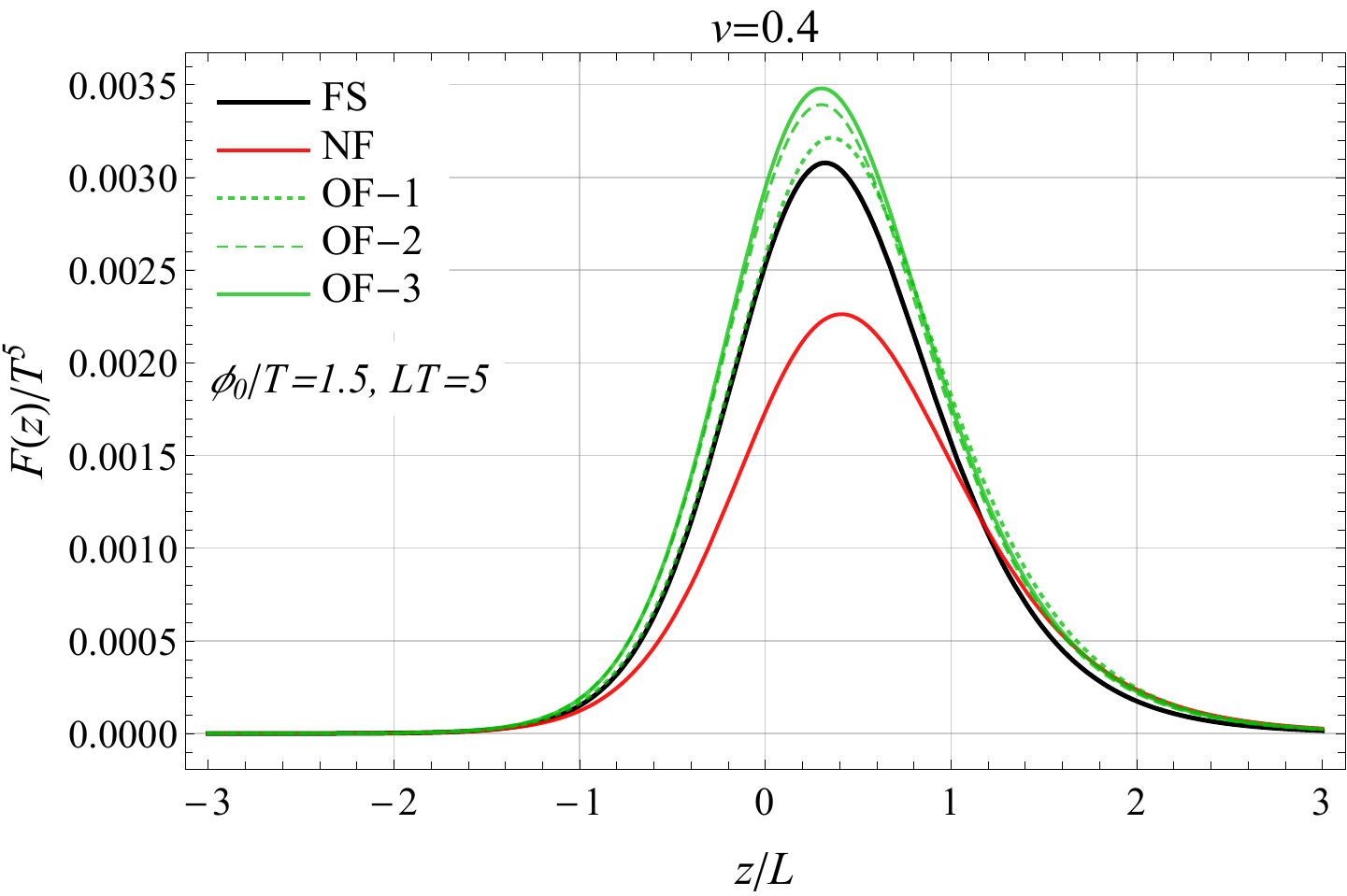}}\\
	\vspace{.5em}
	{\includegraphics[width=.47\textwidth]{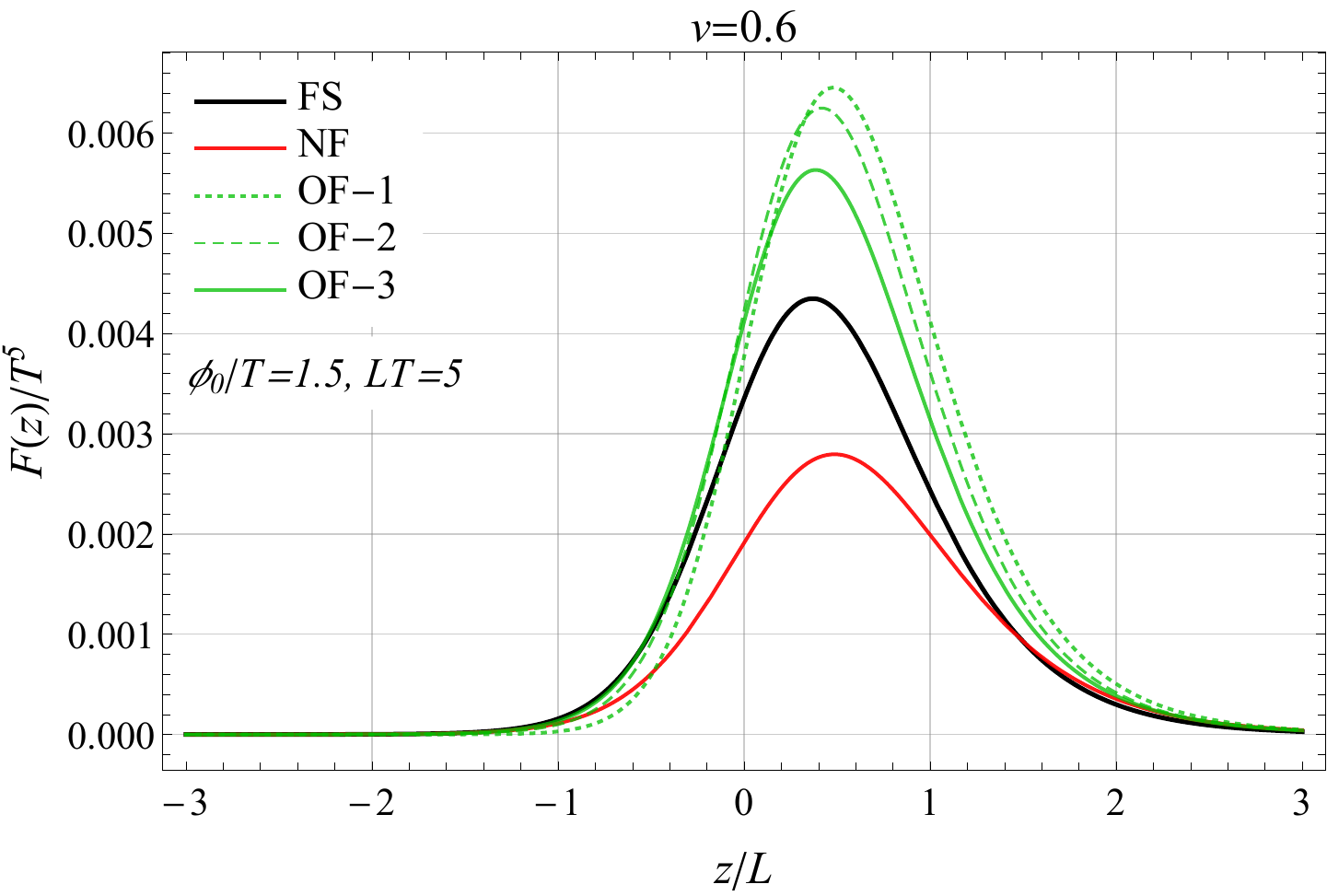}}
	\hfill
	{\includegraphics[width=.47\textwidth]{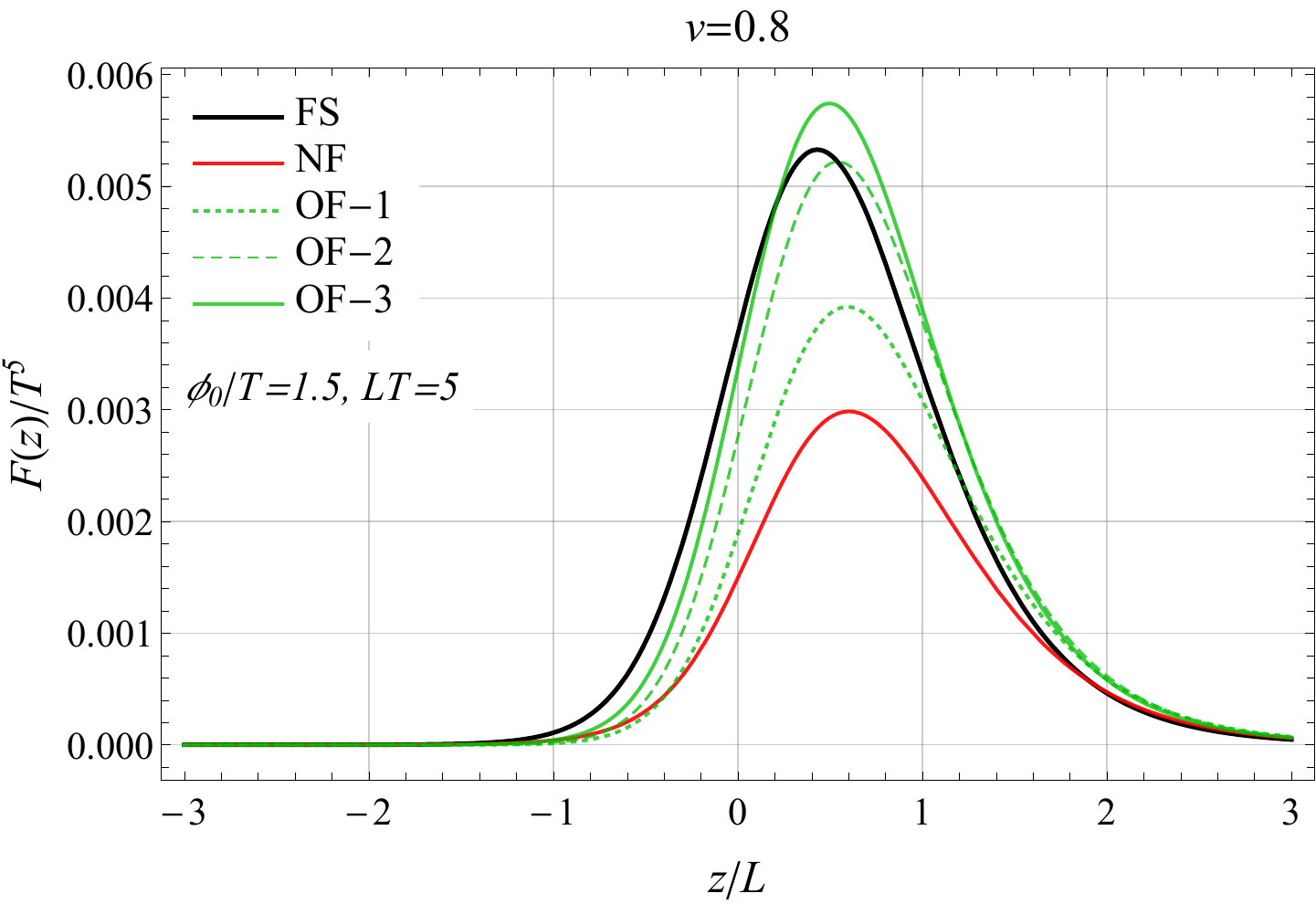}}
	\caption{Friction as a function of the position $z$ when top annihilation and scattering processes are taken into account. The plots correspond to the wall velocities $v  = 0.2, 0.4, 0.6, 0.8$.}\label{fig:friction_ann_scatt}
\end{figure}

The friction as a function of the position $z$ for some benchmark wall velocities is shown in fig.~\ref{fig:friction_ann_scatt}.
Analogously to what we found for the total friction, the results we obtain with our method are only mildly modified by the
scattering contributions. In particular the shape of the friction remains almost unchanged with only minor modifications in the overall
normalization. Similar considerations apply for the shape of the $z$ dependence of the friction in the old and new formalism.
In this case, however, significant changes in the overall normalization are found, as expected from the above discussion on
the total friction.

\begin{figure}
	\centering
	{\includegraphics[width=.24\textwidth]{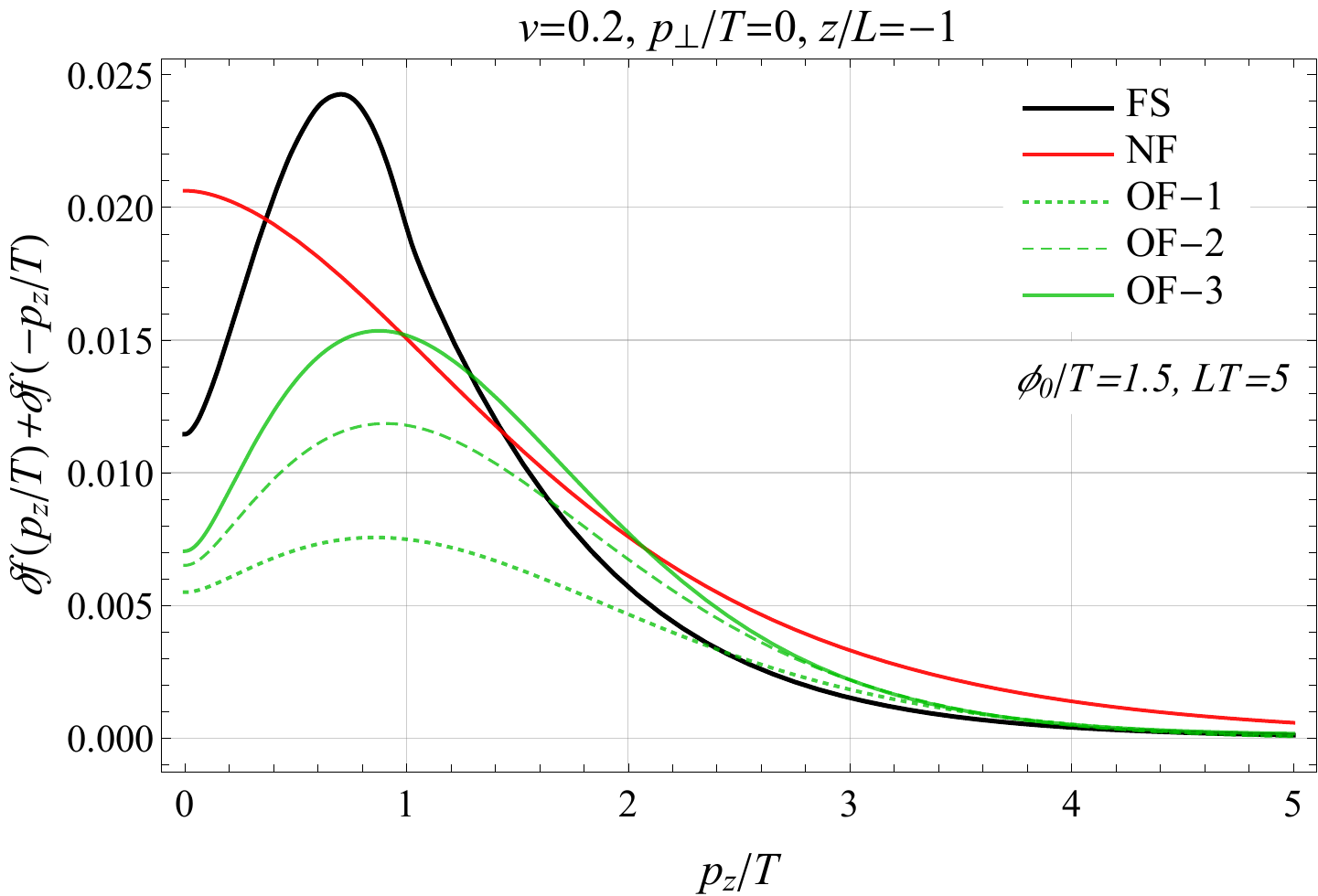}}
	\hfill
	{\includegraphics[width=.24\textwidth]{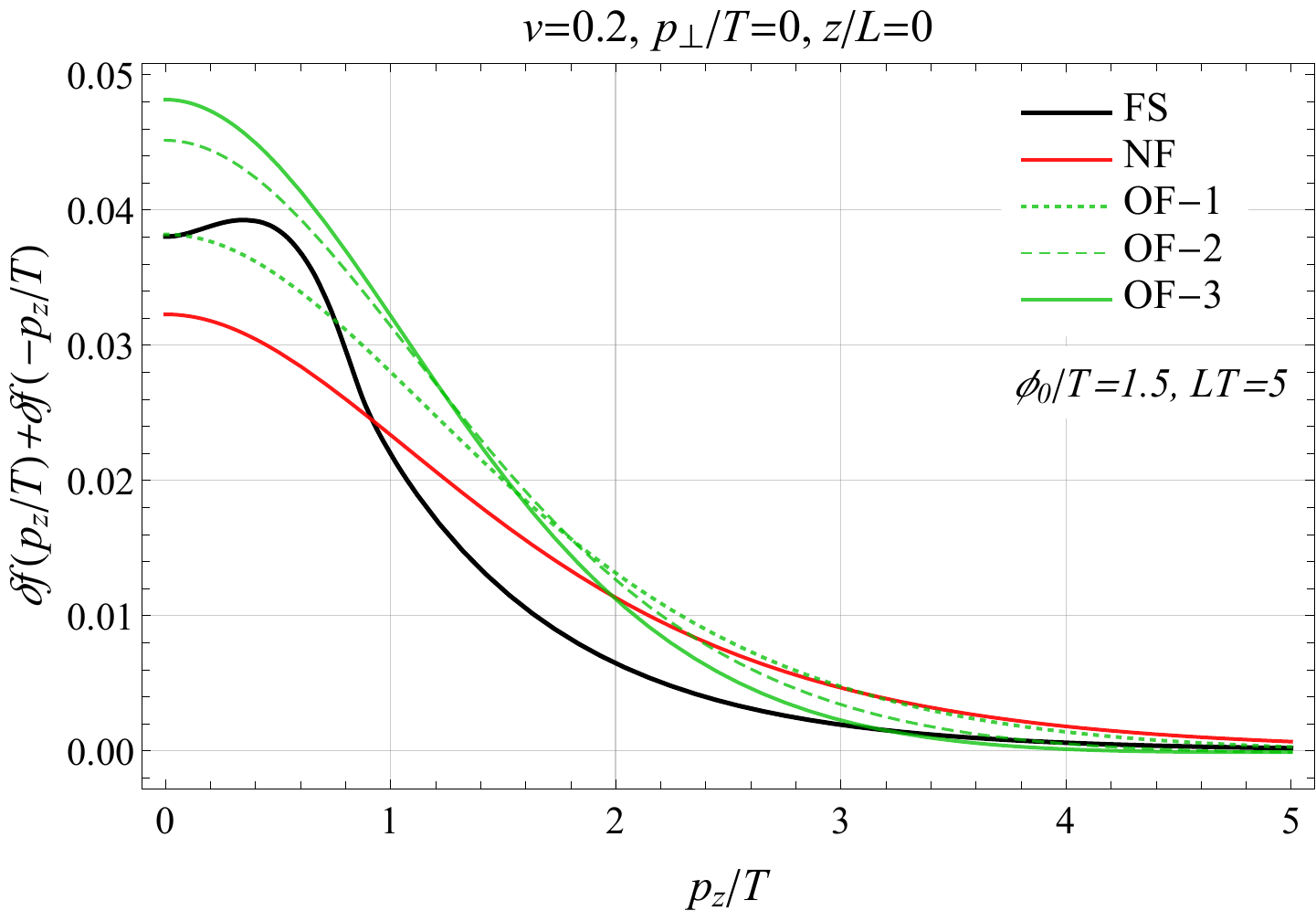}}
	\hfill
	{\includegraphics[width=.24\textwidth]{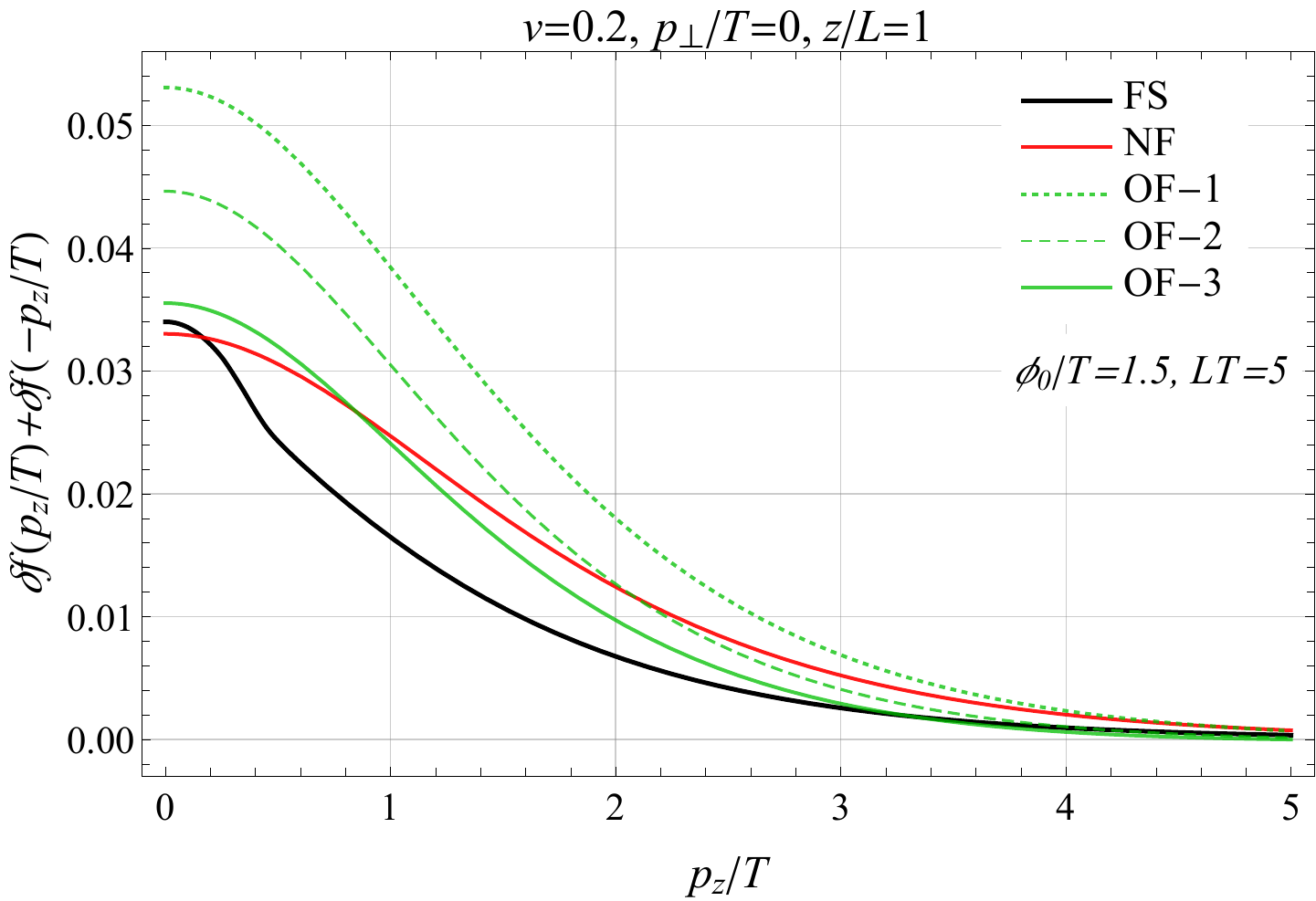}}
	\hfill
	{\includegraphics[width=.24\textwidth]{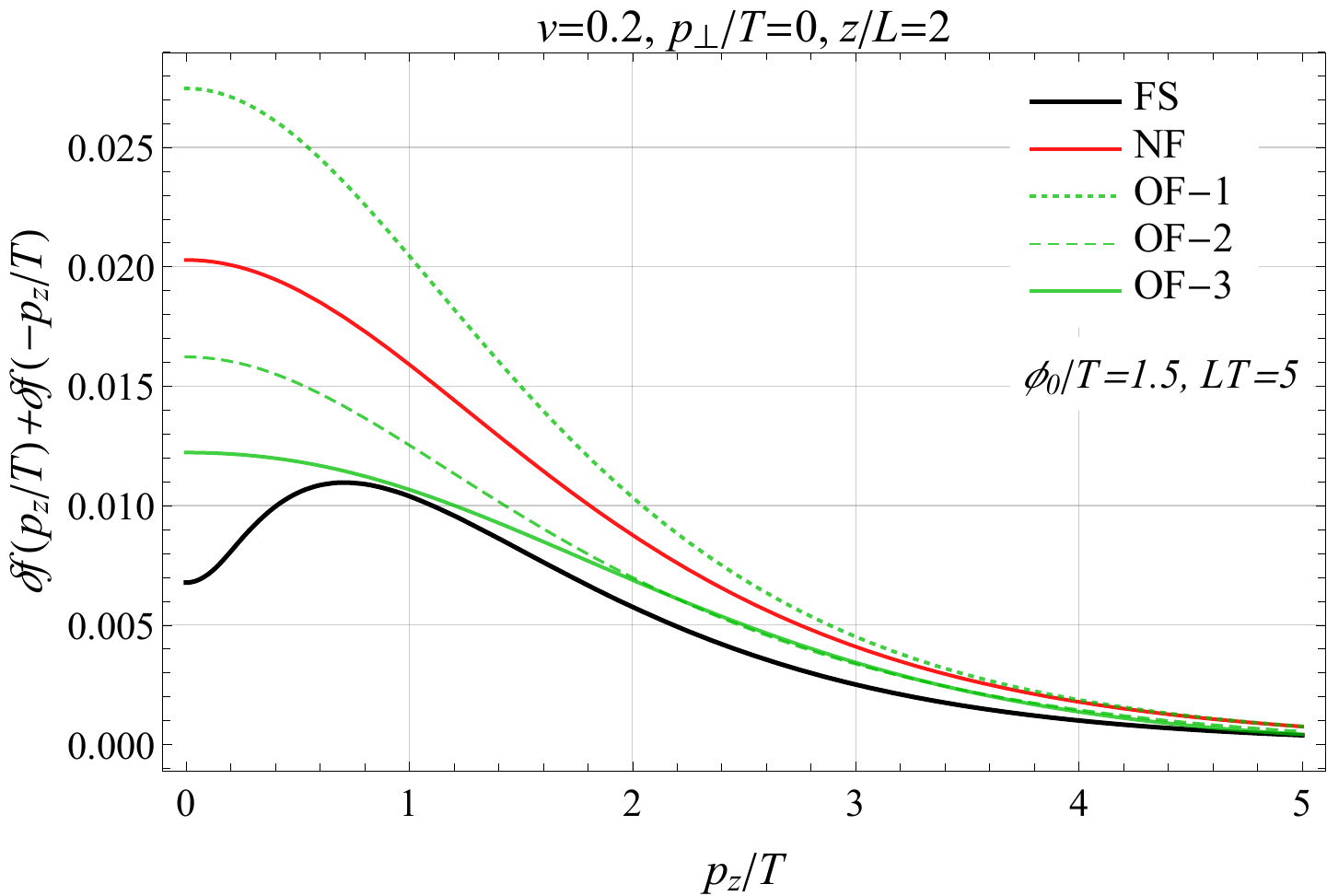}}\\
	\vspace{.5em}
	{\includegraphics[width=.24\textwidth]{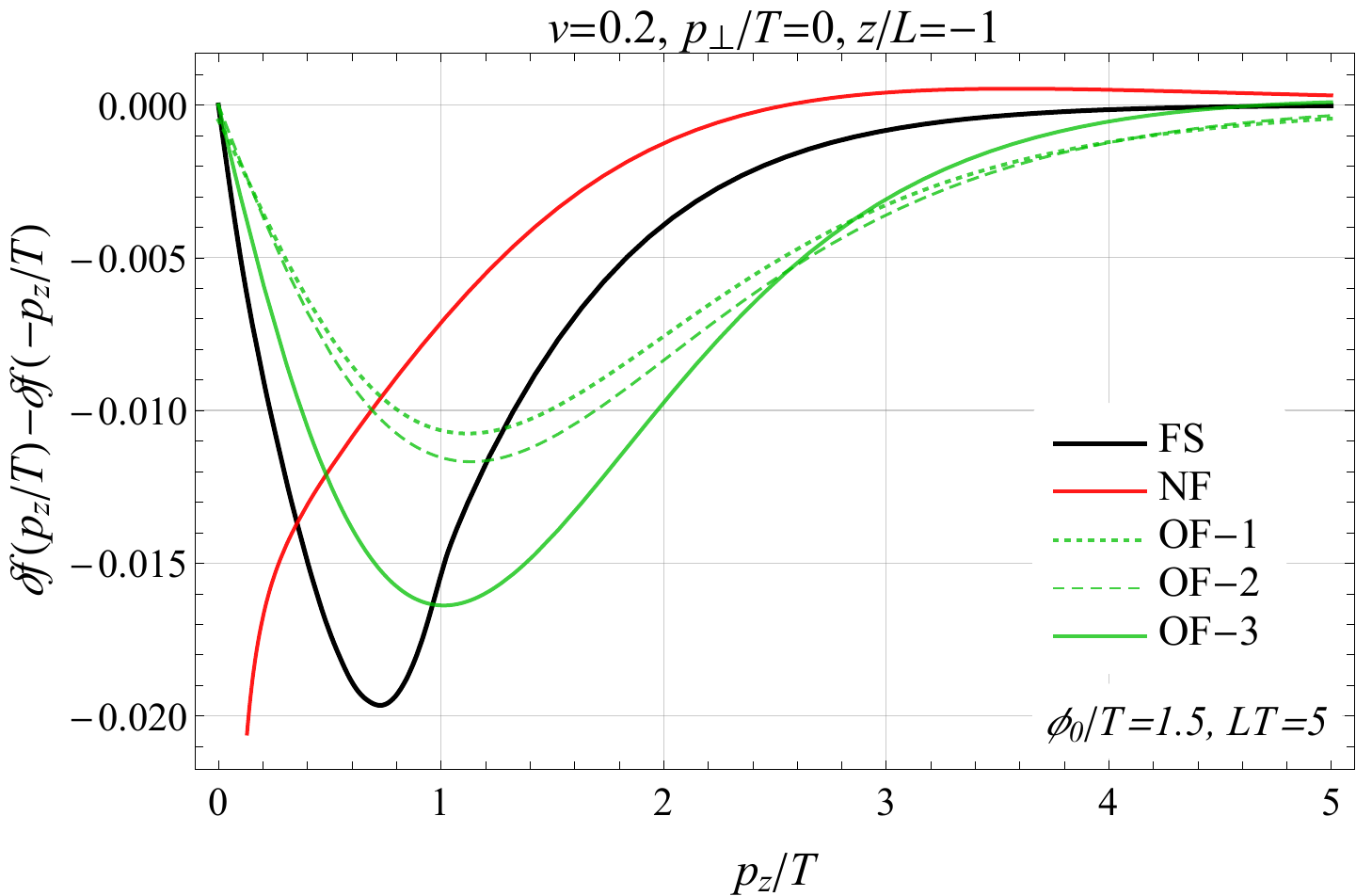}}
	\hfill
	{\includegraphics[width=.24\textwidth]{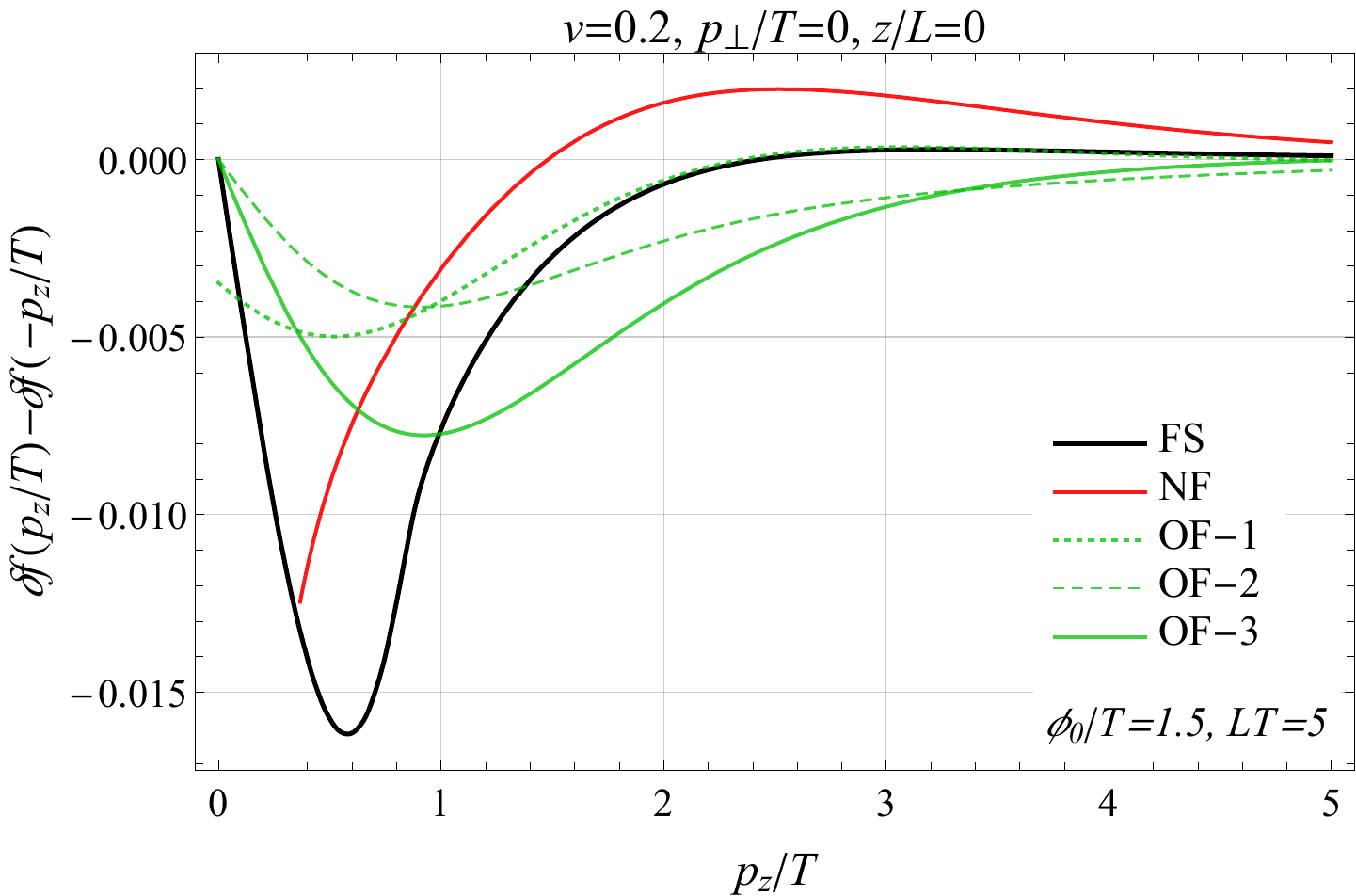}}
	\hfill
	{\includegraphics[width=.24\textwidth]{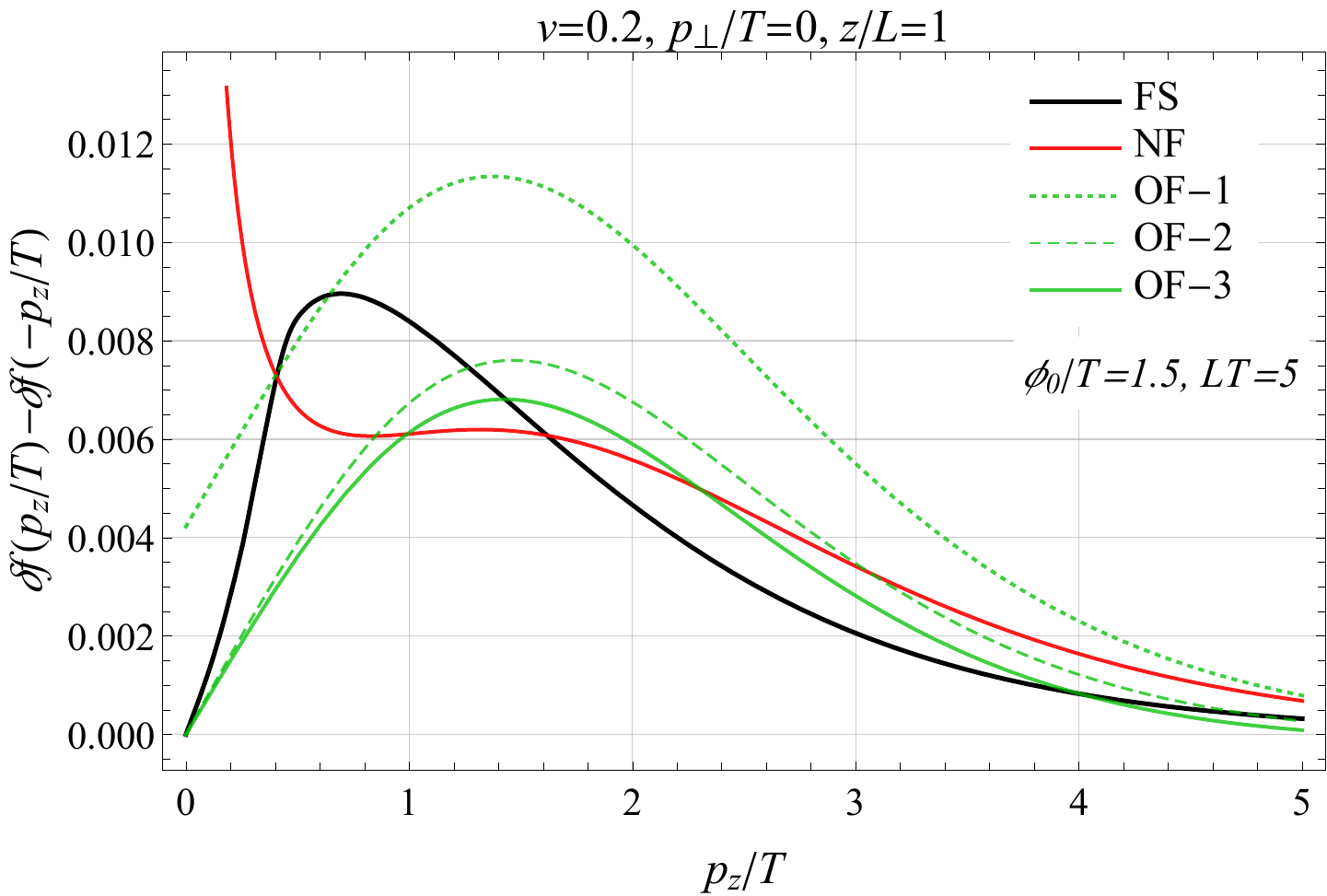}}
	\hfill
	{\includegraphics[width=.24\textwidth]{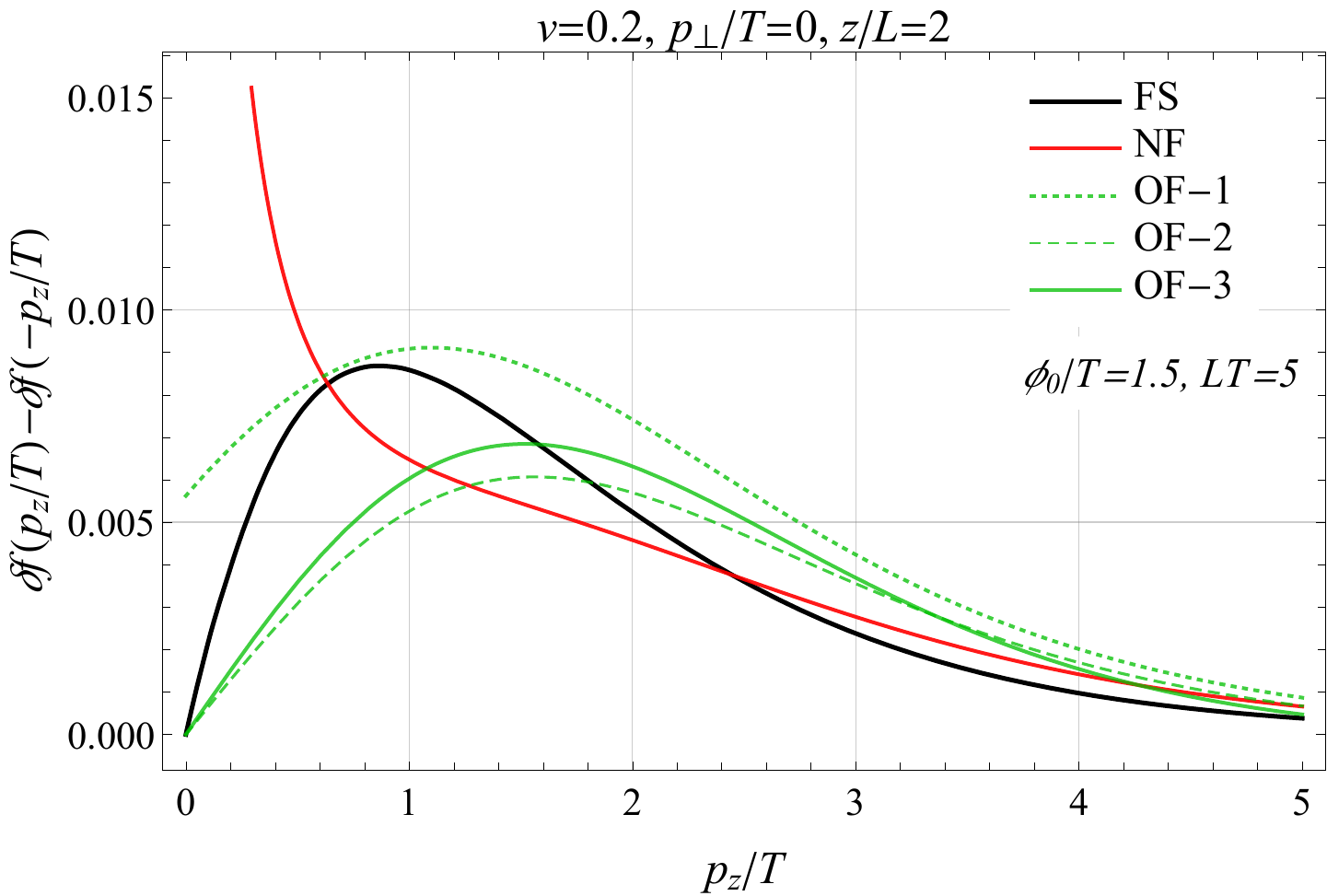}}\\
	\vspace{.5em}
	{\includegraphics[width=.24\textwidth]{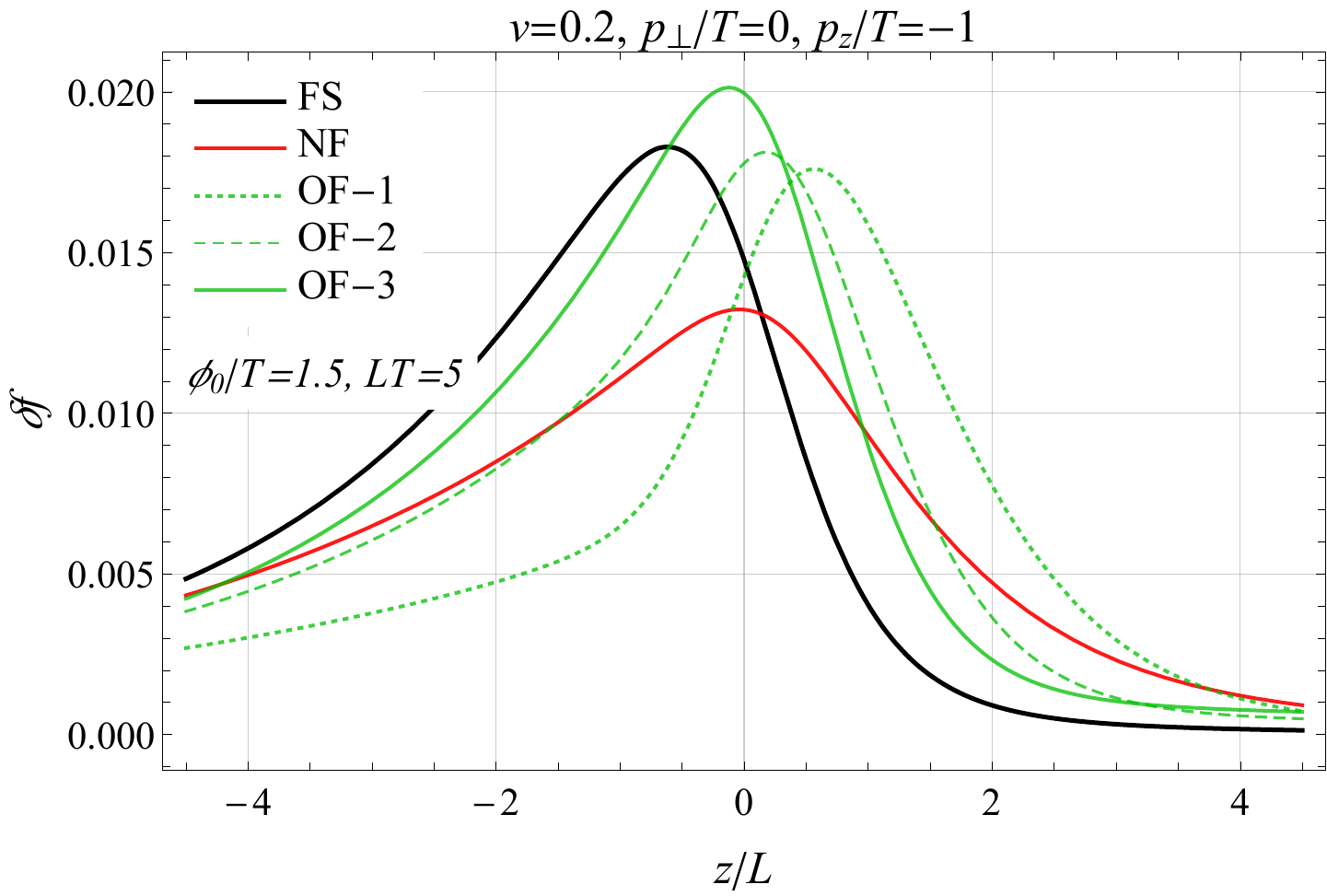}}
	\hfill
	{\includegraphics[width=.24\textwidth]{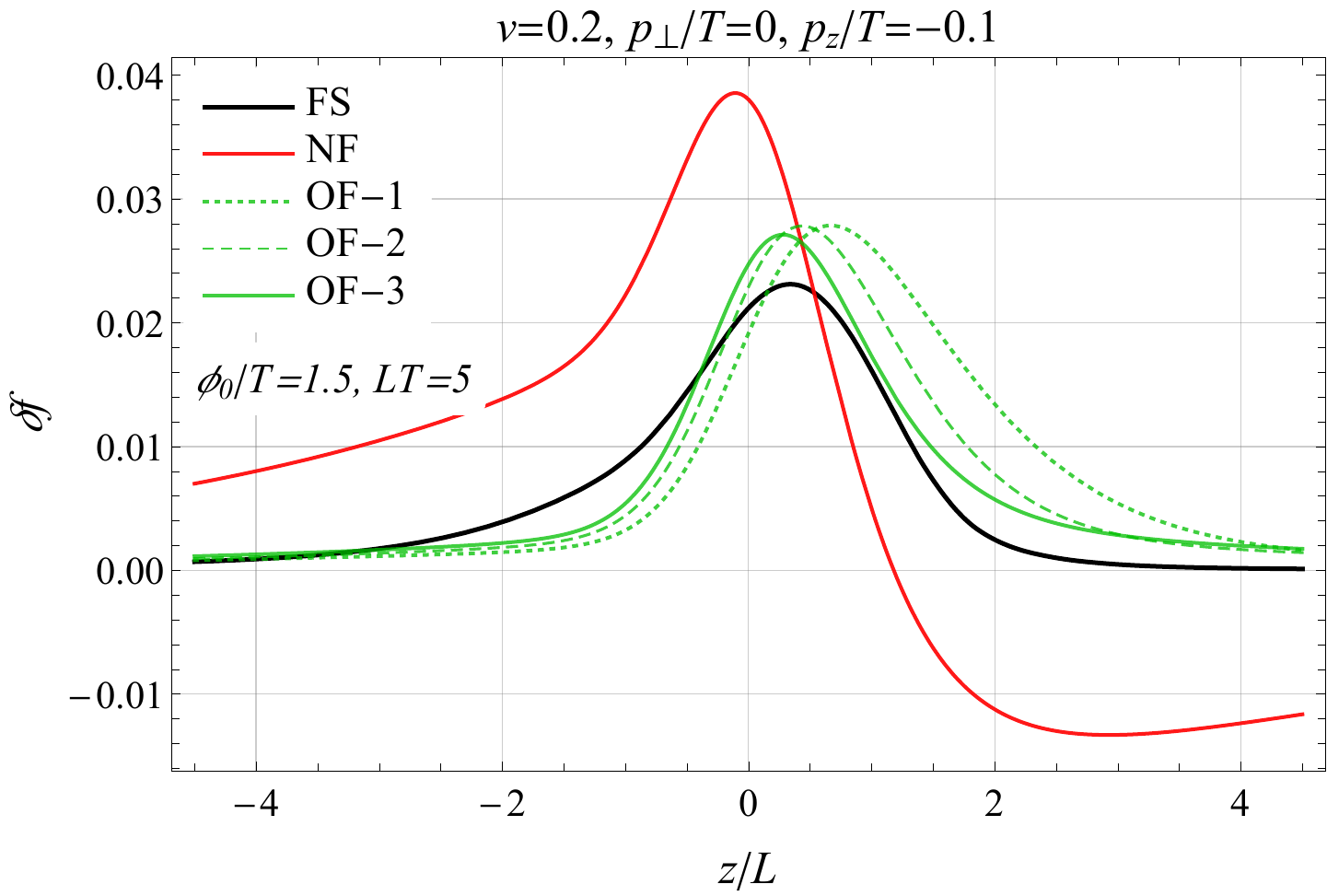}}
	\hfill
	{\includegraphics[width=.24\textwidth]{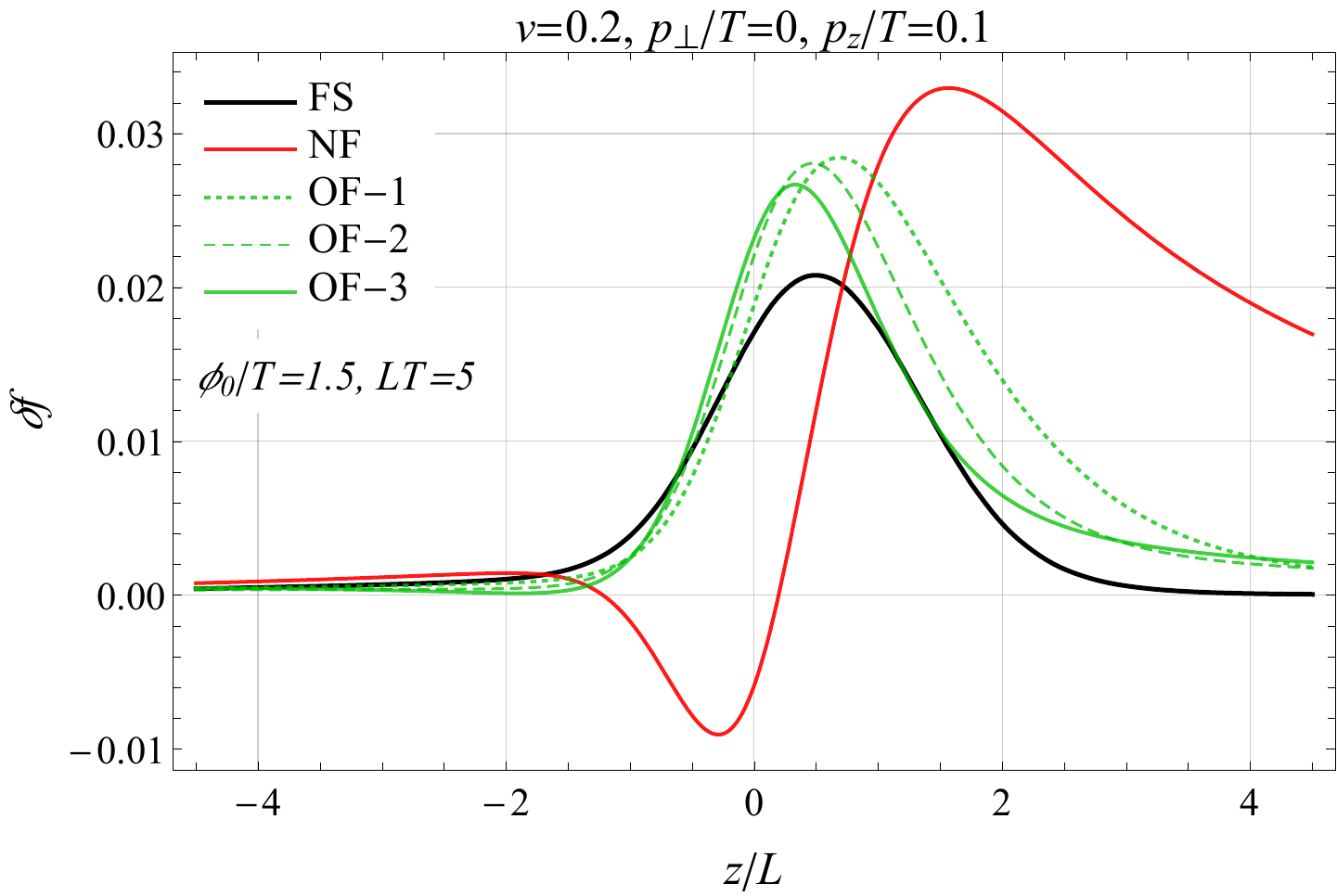}}
	\hfill
	{\includegraphics[width=.24\textwidth]{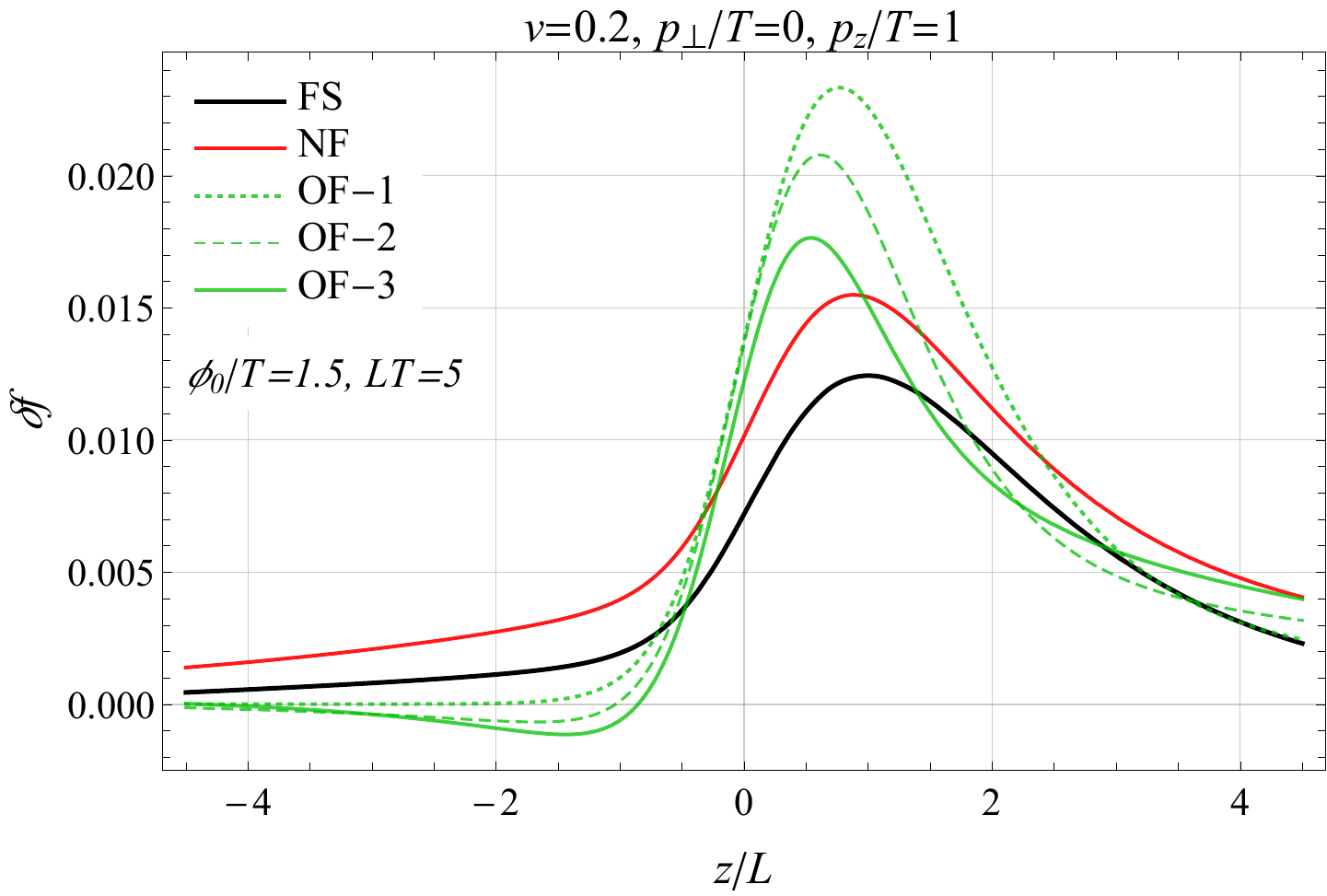}}	
	\caption{Perturbation $\delta f$ for $v = 0.2$ and $p_\bot = 0$. The plots on the first (second) row show the even
	(odd) part of $\delta f$ as a function of $p_z$ for $z/L = -1, 0, 1, 2$. The third row shows plots of $\delta f$ as a function of $z$ for $p_z/T = -1, -0.1, 0.1, 1$.}\label{fig:perturbation_ann_scatt}
\end{figure}

Finally we show in fig.~\ref{fig:perturbation_ann_scatt} the perturbation $\delta f$ for the benchmark velocity $v = 0.2$.
The results for different velocities are qualitatively analogous, the main difference being an overall rescaling with a limited change in shape. The plots show the full solution of the Boltzmann equation we got in our analysis, along with the results obtained applying
the old and new weighted approaches.
Notice that the new formalism does not fully determine the velocity perturbation, whose impact can only be computed
averaging over the momentum through a factorization ansatz~\cite{Cline:2000nw}.
To plot the solution in the new formalism we chose to identify the distribution perturbation with
\begin{equation}
\delta f = - f_v'\, [\mu (z)+\beta \gamma (E - v p_z) \delta \tau] + f_v \frac{E}{p_z} u\,,
\end{equation}
following eq.~(B5) of ref.~\cite{Laurent:2020gpg}.
This identification tends to produce a divergent behavior for small $p_z$, which however has no impact on the determination
of the friction since it is odd in $p_z$.

In the first (second) row of the figure we show the dependence of the even (odd) part of $\delta f$ on the momentum along the $z$
axis, $p_z$. The plots are obtained fixing $p_\bot = 0$, but similar results are found for $p_\bot/T \lesssim 2$
(for larger $p_\bot$ the solution is significantly suppressed and its impact on the domain wall dynamics is subleading).
The plots in the first row show that, at the qualitative level, the old and new formalisms fairly reproduce the overall shape of the even part of the solution, although a somewhat different behavior is found for small $p_z$ ($|p_z/T| \lesssim 0.5$).
This difference is most probably due to the fact that in the weighted approach the
\begin{equation}
- \frac{(m(z)^2)'}{2E} \partial_{p_z} \delta f
\end{equation}
term is neglected. This term, although subleading in most of the kinematic space, dominates close to $p_z = 0$, where the
$(p_z/E) \partial_z$ term vanishes. In spite of the fair qualitative agreement, large quantitative differences are present between the
full solution and the ones obtained with the weighted approaches.

The agreement of the various formalisms in the determination of the odd part of $\delta f$ proves quite poor. In particular,
marked differences are found for $z \lesssim 0$, in which case the high-$p_z$ behavior of the solution is not captured by the old formalism, even including higher-order corrections. A mildly better agreement is found for $z/L \gtrsim 1$. The new formalism tends to
reproduce the correct shape for $p_z/T \gtrsim 1$, but presents large differences for small $p_z$. It must be noticed that the
odd part of $\delta f$ does not contribute to the friction, thus the large differences found among the various solutions do not show up
in the determination of $F(z)$.

On the third row of fig.~\ref{fig:perturbation_ann_scatt} we show $\delta f$ as a function of $z$ for $p_\bot = 0$ and for the
benchmark values $p_z/T = -1, -0.1, 0.1, 1$. The old formalism reproduces the overall qualitative behavior of the solution.
Higher-order terms tend to improve the agreement, although failing to fully reproduce the full solution, especially at the quantitative level.
The new formalism  is in qualitative agreement with the full solution for $|p_z/T| \gtrsim 1$, while completely fails to
reproduce the correct shape for small $p_z$.

\section{Conclusions and Outlook}\label{sec:conclusions}

In this paper we presented for the first time the fully quantitative solution of the Boltzmann equation that describes particle diffusion in the presence of a moving domain wall.
Contrary to the existing approaches, we did not rely on any ansatz nor we imposed any momentum dependence on the non-equlibrium distribution functions.
This clearly represents a necessary step towards a reliable understanding of the bubble wall dynamics. 
Using the friction obtained with the numerical method developed in this work, one can solve the equation of motion of the Higgs profile (or of any other scalar field driving a first order PhT) and extract the velocity of the domain wall as well as the features of its shape, such as the wall thickness. 
These parameters crucially impact on the prospects of any BSM theory to predict interesting cosmological signals, such as a gravitational wave background and the amount of matter-antimatter asymmetry.

We critically compared our results with the ones obtained using the formalisms developed so far in the literature, namely the fluid approximation originally developed in ref.~\cite{Moore:1995si}, its extended version~\cite{Dorsch:2021ubz,Dorsch:2021nje} (we dubbed both approaches `old formalism', following ref.~\cite{Laurent:2020gpg}), and the ``new formalism''~\cite{Laurent:2020gpg}.

To establish our approach, we focused on a slightly simplified set-up in which only the top quark contribution to the DW dynamics is taken into account. Other species can however be included in a straightforward way. We computed numerically the distribution
function for the top species and we obtained the friction $F$ that the plasma exerts on the DW. The latter quantity is shown in fig.~\ref{fig:friction_ann_scatt} as a function of the position $z$ for the three different setups, namely: the old formalism (OF), the new one (NF) and our full solution (FS).
The spatial dependence of the friction is quite similar in the three cases because the overall shape is mainly determined by the source term in the Boltzmann equation, namely $d m^2(z)/dz$. There is however a significant disagreement at the quantitative level,
as can be seen in the right panel of fig.~\ref{fig:friction}, where the integrated friction is plotted as a function of the DW speed $v$.
For small velocities $v \lesssim 0.2$ a good agreement between the new formalism prediction and the full solution is found, while
the old formalism (both original and extended) shows minor differences, of order $10\%$. The agreement significantly worsens at
larger velocities. In this case the new formalism predicts a significantly smaller total friction, reaching a maximum much earlier than
the full solution. The old formalism, on the other hand shows a different qualitative behavior, with a series of peaks related
to the zero eigenvalues of the Liouville operator.\footnote{The presence of peaks depends on the choice of the parameters. They are typically sharper for small wall thickness and weak plasma interaction strength. In other regions of the parameter space, a smooth behavior can be present, as found in ref.~\cite{Dorsch:2021nje}.}
These features, which strongly depend on the order at which one fluid approximation is truncated,
do not seem physical and are not present in the full solution, which shows a completely smooth shape (linear behavior for
$v \lesssim 0.7$ and a flattening for larger velocities).

In fig.~\ref{fig:perturbation_ann_scatt} we also compare the distribution perturbation $\delta f$ for the various approaches.
Although in some kinematic regions a qualitative agreement can be seen, the differences among all approaches are quite strong.
In particular the new formalism shows large differences in the odd part (with respect to $p_z$) of the perturbation. This difference
does not show up in the friction result, since only the even part contributes to $F(z)$.

We add that, as an intermediate step in our procedure, we considered the set-up in which only the annihilation channel for the top
quarks is included in the collision integral, excluding the scattering processes. The friction for the full solution proves remarkably
similar to the one in the complete set-up (see the left panel of fig.~\ref{fig:friction}), apart from the fact that the maximum is reached for
larger DW velocities. The old and new formalisms, on the other hand, show drastically different behavior. In particular the old formalism
predicts sharp peaks connected to the sound speed. The inclusion of higher orders in the fluid approximation does not seem
to achieve convergence in a reliable way. 

As we mentioned, for the purpose of presenting the methodology and setting up the stage for a determination of the velocity of the bubble wall, in the present paper we only considered the top quark contribution to the DW dynamics. The inclusion of the electroweak gauge bosons and of the background species is clearly important to obtain quantitatively reliable predictions.
We leave the investigation of this aspect for future work.

We also exploited another minor simplification in the computation of the collision integrals, ignoring the space dependence of the collisional kernels (see Appendix \ref{app:analytic}), which appears through the top mass in the integrated equilibrium distribution functions. This approximation is also used in the old and new formalisms, and is expected to induce only minor corrections to
the results. Within our approach the full space dependence could be taken into account, at the cost of increasing the computation time.

\section*{Acknowledgments}

We thank J. Kozaczuk and B. Laurent for useful discussions.
L.D.R. has been supported by a fellowship from ``la Caixa'' Foundation (ID 100010434) and from the European Union's Horizon 2020 research and innovation programme under the Marie Sklodowska-Curie Action grant agreement No 847648.
S.D.C. and G.P. were supported in part by the MIUR under contract 2017FMJFMW (PRIN2017). \'A.G.M. has been supported by the Secretariat for Universities and Research of the Ministry of Business and Knowledge of the Government of Catalonia and the European Social Fund and La Caixa through the Becas Postdoctorado Junior Leader (LCF/BQ/PI20/11760032).

\appendix

\section{Evaluation of the collision integrals}\label{app:analytic}

\subsection{The term proportional to $\delta f(p)$}\label{sec:C_deltafp}

We focus, at first, on the term of the collisional integral proportional
to $\delta f(p)$, which, for a single matrix element, reads
\begin{equation}\label{eq:C_bar_int1}
{\cal J}[\delta f]=\frac{-\delta f(p)}{4 N_p E_{p}}\frac{f_{v}(p)}{f_{v}'(p)}\int\!\!\frac{d^{3}{\bf k}\,d^{3}{\bf p'}\,d^{3}{\bf k'}}{(2\pi)^{5}2E_{k}\,2E_{p'}\,2E_{k'}}|{\cal M}|^{2}\delta^{4}(p+k-p'-k'){\cal \,}f_{v}(k)(1\pm f_{v}(p'))(1\pm f_{v}(k'))\,.
\end{equation}
To evaluate the integral it is convenient to change variables through a boost, going to the
plasma frame, in which the Boltzmann distribution is the standard
equilibrium one $f_{v}$. We denote by a bar the momenta in the plasma frame, namely
\begin{equation}
\bar{p}_{0}=\gamma(E_p-vp_{z})\,,\quad\bar{p}_{z}=\gamma(p_{z}-v E_p)\,,\quad\bar{p}_{\bot}=p_{\bot}\,,
\end{equation}
and analogously for $k$, $p'$ and $k'$. We thus get
(notice that the integration measure $d^{3}{\bf p}/E_p$ is invariant
under boost)
\begin{eqnarray}
\overline{\cal J}[\delta f]&=&\frac{-\delta f(p(\bar{p}))}{4 N_p \gamma(E_{\bar{p}}+v\bar{p}_{z})}\frac{f_{0}(\bar{p})}{f_{0}'(\bar{p})}\label{eq:C_bar_int2}\\
&&\times\int\!\!\frac{d^{3}{{\bf \bar k}}\,d^{3}{{\bf \bar p'}}\,d^{3}{{\bf \bar k'}}}{(2\pi)^{5}2E_{\bar{k}}\,2E_{\bar{p}'}\,2E_{\bar{k}'}}|{{\cal M}}|^{2}\delta^{4}(\bar{p}+\bar{k}-\bar{p}'-\bar{k}'){\cal \,}f_{0}(\bar{k})(1\pm f_{0}(\bar{p}'))(1\pm f_{0}(\bar{k}'))\,.\nonumber
\end{eqnarray}

In order to evaluate the integrals, we follow the approach of ref.~\cite{Moore:1995si}, including only leading log contributions.
In this approximation we can also neglect the masses of the particles involved in the scattering. This approximation significantly
simplifies the numerical evaluation, since it removes any explicit dependence on the $z$ coordinate in the integrals.
Closer inspection of the integral appearing in eq.~(\ref{eq:C_bar_int2}) shows that it is invariant under rotation of the
three-momentum components of $\bar p$, thus it is just a function of $E_{\bar p}$.\footnote{Rotation invariance is an immediate
consequence of the fact that the Boltzmann distribution $f_0$ depends only on the energy of the particle, while $|{\cal M}|^2$
is a function of the kinematic invariants (i.e.~the Mandelstam variables).}

The evaluation of the integral can be simplified by exploiting the delta function and the symmetries of the integrand.
In this way one can perform analytically five of the nine integrals. An efficient parametrization for performing the integration
is presented in ref.~\cite{Arnold:2003zc}.

In the leading log approximation, only t-channel and u-channel scattering amplitudes are relevant (see table~\ref{tab:amplitudes}).
So we can focus on these two types of contributions and neglect s-channel processes (and interference terms).

\begin{table}
\centering
\begin{tabular}{c|c}
process & $|{\cal M}|^2$\\
\hline
\rule{0pt}{1.75em}$t \bar t \to gg$ & $\displaystyle \frac{128}{3} g_s^4 \left[ \frac{ut}{(t - m_q^2)^2} +  \frac{ut}{(u- m_q^2)^2} \right ]$\\
\rule{0pt}{1.75em}$tg \to tg$ & $\displaystyle- \frac{128}{3} g_s^4 \frac{su}{(u-m_q^2)^2} + 96 g_s^4 \frac{s^2 + u^2}{(t - m_g^2)^2}$\\
\rule{0pt}{1.75em}$tq \to tq$ & $\displaystyle160 g_s^4 \frac{s^2 + u^2}{(t - m_g^2)^2}$
\end{tabular}
\caption{Amplitudes for the scattering processes relevant for the top quark in the leading log approximation. In the $t q \to t q$ process we summed over all massless quarks and antiquarks.}\label{tab:amplitudes}
\end{table}

\subsubsection*{t-channel parametrization}

We start by considering amplitudes coming from t-channel diagrams.
The integration over $d^{3}\bar{{\bf k'}}$ can be easily performed exploiting
the $\delta$-function. The remaining integrals can be handled through a change of variables.
As in ref.~\cite{Arnold:2003zc}, we introduce
the three-momentum $\mathbf{q}\equiv\mathbf{\bar{p}}'-\mathbf{\bar{p}}=\bar{\mathbf{k}}-\mathbf{\bar{k}}'$.
Rotational invariance allows us to trivially integrate on the orientation
of $\mathbf{q}$. Fixing $\mathbf{q}$ to be along a $z'$ axis, we
can express the orientation of the $\mathbf{\bar{p}}$ and $\mathbf{\bar{k}}$
momenta in terms of the polar angles $\theta_{\bar{p}q}$ and $\theta_{\bar{k}q}$ and
the azimuthal angle $\phi$ between the $\mathbf{\bar{p}}$-$\mathbf{q}$
and the $\mathbf{\bar{k}}$-$\mathbf{q}$ plane.

The remaining delta function can be handled by introducing an additional variable $\omega$
linked to the $t$ Mandelstam variable as $t \equiv \omega^2 - q^2$, where $q \equiv |{\bf q}|$. In this way the
integrations on the angles $\theta_{\bar{k}q}$ and $\theta_{\bar{p}q}$ can be performed analytically
and one is left with the final expression for the integral on the second line of eq.~(\ref{eq:C_bar_int2}):
\begin{equation}
{\cal K}=\frac{1}{8(2\pi)^{4}E_{\bar{p}}}\int\limits _{-E_{\bar{p}}}^{+\infty}d\omega\int\limits _{|\omega|}^{\omega+2E_{\bar{p}}}dq\int\limits _{\frac{q+\omega}{2}}^{+\infty}d E_{\bar{k}}\int\limits _{0}^{2\pi}d\phi\,|{{\cal M}}|^{2}f_{0}(\bar{k})(1\pm f_{0}(\bar{p}'))(1\pm f_{0}(\bar{k}'))\,.
\end{equation}

As alternative parametrization, which can help in the numerical integration and in studying the behavior of the integral,
one can define
\begin{equation}
\chi_{\pm}\equiv q\pm\omega\,,
\end{equation}
in terms of which
\begin{equation}
\int\limits _{-E_{\bar{p}}}^{+\infty}d\omega\int\limits _{|\omega|}^{\omega+2E_{\bar{p}}}dq\int\limits _{\frac{q+\omega}{2}}^{+\infty}d E_{\bar{k}}\quad\rightarrow\quad\frac{1}{2}\int\limits _{0}^{+\infty}d\chi_{+}\int\limits _{0}^{2 E_{\bar{p}}}d\chi_{-}\int\limits _{\chi_{+}/2}^{\infty}d E_{\bar{k}}\,.
\end{equation}
The $\chi_{\pm}$ parametrization can be also useful to leave as last integration the one on $E_{\bar{k}}$:
\begin{equation}
\frac{1}{2}\int\limits _{0}^{+\infty}d\chi_{+}\int\limits _{0}^{2 E_{\bar{p}}}d\chi_{-}\int\limits _{\chi_{+}/2}^{\infty}d E_{\bar{k}}\quad\rightarrow\quad\frac{1}{2}\int\limits _{0}^{\infty}d E_{\bar{k}}\int\limits _{0}^{2 E_{\bar{k}}}d\chi_{+}\int\limits _{0}^{2 E_{\bar{p}}}d\chi_{-}\,.
\end{equation}
This choice of integration order clearly shows the symmetric role of $E_{\bar{p}}$ and $E_{\bar{k}}$ in the collisional integral.

The expressions for the $s$ and $u$ Mandelstam variables as a function of $\omega$, $q$ and $E_{\bar k}$ are given by
\begin{eqnarray}
s & = & -\frac{t}{2q^{2}}\left\{ \left[(2 E_{\bar{p}}+\omega)(2 E_{\bar{k}}-\omega)+q^{2}\right]-\cos\phi\sqrt{(4 E_{\bar{p}}(E_{\bar{p}}+\omega)+t)(4 E_{\bar{k}}(E_{\bar{k}}-\omega)+t)}\right\}\,,\hspace{1.5em}\\
t & = & \omega^{2}-q^{2}\,,\\
u & = & -t-s\,,
\end{eqnarray}
while the relative angles between the three-momenta are given in ref.~\cite{Arnold:2003zc} (see Appendix A.2, eqs.~(A21a)-(A21e)),
among which
\begin{equation}
\cos\theta_{pq}=\frac{\omega}{q}+\frac{t}{2 E_{\bar{p}}q}\,,\qquad\cos\theta_{kq}=\frac{\omega}{q}-\frac{t}{2E_{\bar{k}}q}\,.
\end{equation}

\subsubsection*{u-channel parametrization}

Analogous formulae can be found for the u-channel parametrization,
by exchanging $\bar{\mathbf{p}}'$ and $\bar{\mathbf{k}'}$ in the
t-channel parametrization. In this way the integral becomes
\begin{equation}
{\cal K}=\frac{1}{8(2\pi)^{4} E_{\bar{p}}}\int\limits _{- E_{\bar{p}}}^{+\infty}d\omega\int\limits _{|\omega|}^{\omega+2 E_{\bar{p}}}dq\int\limits _{\frac{q+\omega}{2}}^{+\infty}d E_{\bar{k}}\int\limits _{0}^{2\pi}d\phi\,|{{\cal M}}|^{2}f_{0}(\bar{k})(1\pm f_{0}(\bar{p}'))(1\pm f_{0}(\bar{k}'))\,,
\end{equation}
with $\mathbf{q}\equiv\mathbf{\bar{k}}'-\mathbf{\bar{p}}=\mathbf{\bar{k}}-\mathbf{\bar{p}}'$
and
\begin{equation}
\omega=E_{\bar{k}'}- E_{\bar{p}}=E_{\bar{k}}- E_{\bar{p}'}\,.
\end{equation}
 The expressions for the $s$ and $u$ Mandelstam variables are given
by
\begin{eqnarray}
s & = & -\frac{u}{2q^{2}}\left\{ \left[(2 E_{\bar{p}}+\omega)(2 E_{\bar{k}}-\omega)+q^{2}\right]-\cos\phi\sqrt{(4 E_{\bar{p}}(E_{\bar{p}}+\omega)+u)(4E_{\bar{k}}(E_{\bar{k}}-\omega)+u)}\right\}\,,\hspace{1.5em}\\
u & = & \omega^{2}-q^{2}\,,\\
t & = & -u-s\,,
\end{eqnarray}
 while
\begin{equation}
\cos\theta_{pq}=\frac{\omega}{q}+\frac{u}{2E_{\bar{p}}q}\,,\qquad\cos\theta_{kq}=\frac{\omega}{q}-\frac{u}{2 E_{\bar{k}}q}\,.
\end{equation}

\subsubsection*{Structure of the contribution}

From the above formulae we can easily infer the global structure of
the collisional term proportional to $\delta f(p)$. The quantity
(we consider the $t$-channel parametrization for definiteness)
\begin{equation}
{\cal K}=\frac{1}{8(2\pi)^{4} E_{\bar{p}}}\int\limits _{-E_{\bar{p}}}^{+\infty}d\omega\int\limits _{|\omega|}^{\omega+2 E_{\bar{p}}}dq\int\limits _{\frac{q+\omega}{2}}^{+\infty}d E_{\bar{k}}\int\limits _{0}^{2\pi}d\phi\,|{{\cal M}}|^{2}f_{0}(\bar{k})(1\pm f_{0}(\bar{p}'))(1\pm f_{0}(\bar{k}'))\,,
\end{equation}
only depends on $E_{\bar{p}}$, as we already anticipated.
Therefore we get
\begin{equation}
\overline{\cal J}[\delta f]=- \frac{1}{4N_p}\frac{\delta f(p(\bar{p}))}{\gamma(E_{\bar{p}}+v\bar{p}_{z})}\frac{f_{0}(\bar{p})}{f_{0}'(\bar{p})}{\cal K}[E_{\bar{p}}]\,.
\end{equation}

We can now go back to the wall frame, obtaining
\begin{equation}
{\cal J}[\delta f]=-\frac{1}{4N_p}\frac{\delta f(p)}{E_p}\frac{f_{v}(p)}{f_{v}'(p)}{\cal K}[\gamma(E_p-vp_{z})]
= \frac{1}{4N_p} \frac{\delta f(p)}{E_p}\left(1\pm e^{-\beta\gamma(E_p-vp_{z})}\right){\cal K}[\gamma(E_p-vp_{z})]\,.
\end{equation}
Notice that the massless-limit approximation introduced a small `mismatch' in this expression, since we chose $f_{v}(p)$
in the prefactors to have the full mass dependence (from the definition of $E_p$). In the approach to the solution via the use
of weights, instead, $f_{v}$ is treated in the massless limit for all the factors in the collisional integrals.
This problem could be solved by also considering the massive form for all the $f_{v}$ factors inside the collisional integral.
This however is computationally more demanding, since it introduces an explicit $z$ dependence in the integrand, so that
the kernel should be evaluated also as a function of $z$.\footnote{Notice that a full treatment would also need a redefinition
of the matrix element $|{\cal M}|^{2}$ and of the integration boundaries.}

The numerical analysis shows a behavior
\begin{equation}
{\cal K}(E_{\bar{p}}) \sim \log E_{\bar{p}}+const
\end{equation}
which, as expected, has a logarithmic divergence for mass and thermal mass going to zero. We can thus infer the rough
behavior (at least for small $v$)
\begin{equation}
{\cal J}[\delta f] \sim \delta f(p)\frac{\log E_p + const}{E_p}\,.\label{eq:C_behavior}
\end{equation}

\subsection{The terms $\langle \delta f\rangle$}\label{sec:bk_deltafp}

The second ingredient we need in order to compute the collision integrals is the determination of the terms $\langle\delta f\rangle$
in which the perturbation appears under the integral sign. The generic structure of the term that depends on $\delta f(k)$ is
\begin{equation}\label{eq:CI_dfk}
\langle\delta f(k)\rangle = \frac{- f_v(p)}{4N_p E_p}\int\!\!\frac{d^3{\bf k}\,d^3{\bf p'}\,d^3{\bf k'}}{(2\pi)^5\,2E_k\,2E_{p'}\,2E_{k'}}|{\cal M}|^2\delta^4(p+k-p'-k')f_v(k)(1\pm f_v(p'))(1\pm f_v(k'))\frac{\delta f(k)}{f'_v(k)}\,,
\end{equation}
and analogous expressions are valid for the $\delta f(p')$ and $\delta f(k')$ contributions.

The above integral can in principle be evaluated using the same manipulations we described in section~\ref{sec:C_deltafp}.
However, the integrand, due to the $\delta f(k)$ factor, is not rotationally invariant, and an additional integration over the
direction of $\bf k$ with respect to the $z$ axis remains.
The final result is (also in this section we treat all the particles as massless)
\begin{equation}
\langle\delta f(k)\rangle = -\frac{1}{32N_p}\frac{f_0(p)}{(2\pi)^5 E_p} {\cal I}[p_\bot, p_z, z]\,,
\end{equation}
where ${\cal I}$, written as a function of the $p$ momentum in the plasma frame, reads
\begin{equation}
{\cal I} = \int_{-\bar E_p}^{+\infty}d\omega\int_{|\omega|}^{\omega+2 E_{\bar p}}dq\int_{\frac{q+\omega}{2}}^{+\infty}d E_{\bar k}\int_0^{2\pi} d\phi\int_0^{2\pi}d\phi_k|{\cal M}|^2f_0(\bar k)(1\pm f_0(\bar p'))(1\pm f_0(\bar k'))\frac{\delta f(\bar k)}{f'_0(\bar k)}\,,
\end{equation}
in which $\phi_k$ denotes the angle between the vector ${\bf k}$ and the plane where ${\bf p}$ and $\hat z$, the direction along which the wall moves, lie.
The integral ${\cal I}$ depends on the three variables $p_\bot$, $p_z$ and $z$, and requires five numerical integrations.
Therefore its evaluation on a fine grid, as required in our numerical approach, is quite cumbersome.

An alternative procedure to manipulate the integral can be used to reduce the number of numerical integrations.
This can be done by performing the integration over ${\bf p}'$ and ${\bf k}'$ in eq.~(\ref{eq:CI_dfk}) and leaving the integral over
${\bf k}$ as a last step. In this way the expression for $\langle \delta f(k)\rangle$ can be brought to the form
\begin{equation}
\langle\delta f(k)\rangle = -\frac{f_v(p)}{4N_p E_p} \int\frac{d^3{\bf k}}{2 E_k}{\cal K}_1\, f_v(k)\frac{\delta f(k)}{f'_v(k)}\,,
\end{equation}
where
\begin{equation}
{\cal K}_1 = \frac{1}{(2\pi)^5}\int\frac{d^3{\bf k'}\,d^3{\bf p}}{2 E_{p'}\,2 E_{k'}}|{\cal M}|^2(1\pm f_v(p'))(1\pm f_v(k'))\delta^4(p+k-p'-k')\,.
\end{equation}
Since ${\cal K}_1$ is a Lorentz scalar, it will be a function of the only Lorentz scalars that can be obtained from the four vectors $p^\mu$, $k^\mu$ and the plasma velocity $u^\mu$, namely, $u^\mu p_\mu$, $u^\mu k_\mu$ and $p^\mu k_\mu$.
These quantities are related, respectively, to the energies $E_{\bar p}$ and $E_{\bar k}$ of the incoming particles in the plasma reference frame
and to the angle $\theta_{\bar p \bar k}$ between the momenta $\bar p$ and $\bar k$. Putting everything together we find
\begin{equation}
\langle\delta f(k)\rangle = -\frac{f_v(p)}{4N_p E_p} \int \frac{d^3{\mathbf{\bar{k}}}}{2 E_{\bar{k}}}{\cal K}_1(E_{\bar{p}}, E_{\bar{k}}, \theta_{\bar{p}\bar{k}})\,f_0(\bar{k})\frac{\delta f(k_\perp,\gamma(\bar{k}_z+v E_{\bar{k}}),z)}{(-f'_0(\bar{k}))}\,.
\end{equation}
which can be rewritten as
\begin{equation}
\langle\delta f(k)\rangle= -\frac{f_v(p)}{4N_p E_p} \frac{1}{2}  \int_0^{\infty}\!\!\! E_{\bar{k}}\, d E_{\bar{k}}\int_{-1}^1\!\!\!d\cos\theta_{\bar{p}\bar{k}}\,{\cal K}_1(E_{\bar{p}}, E_{\bar{k}} ,\theta_{\bar{p}\bar{k}})\int_0^{2\pi}\!\!\!\!d\phi_{\bar{k}}\,f_0(\bar{k})\frac{\delta f(k_\perp,\gamma(\bar{k}_z+v E_{\bar{k}}),z)}{(-f'_0(\bar{k}))}\,.
\end{equation}

The collision integral for the scattering processes includes an additional set contributions in which $\delta f(p')$
or $\delta f(k')$ appears. In analogy to the previous case, for the $\delta f(p')$ terms, we can first perform the integrals over ${\bf k}$ and ${\bf k'}$,
obtaining the following expression
\begin{equation}
\langle\delta f(p')\rangle= \frac{- f_v(p)}{4N_p E_p} \int\frac{d^3{\bf p'}}{2 E_{p'}}{\cal K}_2(E_p, E_{p'},\theta_{pp'})(1\pm f_v( p'))\frac{\delta f(p')}{f_v'(p')}\,,
\end{equation}
which can also be rewritten as
\begin{eqnarray}
\langle\delta f(p')\rangle &=& \frac{-f_v(p)}{4N_p E_p} \frac{1}{2}\int_0^\infty E_{\bar{p}'}\,d E_{\bar{p}'}\int_{-1}^1 d\cos\theta_{\bar{p}\bar{p}'}{\cal K}_2(E_{\bar{p}}, E_{\bar{p}'},\theta_{\bar{p}\bar{p}'})\times\nonumber\\
&& \hspace{5em}\times\int_0^{2\pi} d\phi_{\bar{p}'}(1 \pm f_0(\bar{p}'))\frac{\delta f(p'_\perp,\gamma(\bar{p}'_z+v E_{\bar{p}'}),z)}{(-f'_0(\bar{p}'))}\,.
\end{eqnarray}
The contributions from  $\delta f(k')$ can be treated in an analogous way.

\subsubsection{Evaluation of the ${\cal K}_1$ kernel}

The evaluation of the kernel ${\cal K}_1$ can be performed as in ref.~\cite{DeGroot:1980dk}. As a first step we perform the integration over ${\bf k}'$ exploiting the Dirac delta:
\begin{equation}
{\cal K}_1=\frac{1}{(2\pi)^5}\int \frac{d^3{\bf p'}}{2 E_{p'}}\frac{1}{2 E_{k'}}|{\cal M}|^2(1\pm f_0(u^\mu p'_\mu))(1\pm f_0(u^\mu k'_\mu))\delta(E_p+E_k-E_{p'}-E_{k'}).
\end{equation}
Notice that in the above expression we expressed the energies $E_{p'}$ and $E_{k'}$ in the Lorentz-invariant form
$u^\mu p'_\mu$ and $u^\mu k'_\mu$. As we will see, this is useful to keep track of the changes of reference frame.

As a second step, we rewrite the Dirac delta (in the center-of-mass (COM) frame) as
\begin{equation}
\delta(E_p+E_k-E_{p'}-E_{k'})=\delta(\sqrt{s}-2E_{p'}) = \frac{1}{2}\delta\left(\frac{1}{2}\sqrt{s}-E_{p'}\right)\,,
\end{equation}
with $s = (p + k)^2$ the usual Mandelstam variable.
The integration over ${\bf p}'$ can be performed by rewriting $d^3{\bf p}' = E_{p'}^2\, d E_{p'}\, d\cos\theta\, d\phi$,
where $\theta$ is the angle between ${\bf p}'$ and ${\bf p}$ in the COM frame of the scattering process:
\begin{equation}
{\cal K}_1=\frac{1}{(2\pi)^5}\frac{1}{8}\int_{-1}^1d\cos\theta\int_0^{2\pi}d\phi|{\cal M}|^2(1\pm f_0(u^\mu p'_\mu))(1\pm f_0(u^\mu k'_\mu))\,.
\end{equation}

As a last step we need to compute $u^\mu p'_\mu$ and $u^\mu k'_\mu$ in the COM frame.
We conveniently choose the orientation of the COM frame axes such that $u^y=0$ leading to
\begin{equation}
u^\mu p'_\mu = u^0\frac{\sqrt{s}}{2} - u^x\frac{\sqrt{s}}{2}\sin\theta\cos\phi - u^z\frac{\sqrt{s}}{2}\cos\theta 
\end{equation}
We then introduce the four-vectors $P^\mu$ and $Q^\mu$ defined as
\begin{equation}
P^\mu = p^\mu+k^\mu\,,\qquad\quad
Q^\mu =p^\mu-k^\mu\,.
\end{equation}
In the COM frame we find that
\begin{equation}
P^\mu = \left(\begin{array}{c}\sqrt{s}\\0\end{array}\right)\,,
\qquad\quad
Q^\mu = \left(\begin{array}{c}0\\ \mathbf{Q}\end{array}\right)\,.
\end{equation}
We can get a further simplification by choosing the frame such that ${\bf Q}$ lies along the $z$ axis.
Since, in the massless case, $| {\bf P} |=|{\bf Q}|=\sqrt{s}$, the vectors $P^\mu/\sqrt{s}$ and $Q^\mu/\sqrt{s}$ coincide with the versors
along the first and fourth Minkowski directions.

The $u^0$ and $u^z$ components can be easily computed in terms of the momenta of the particles in the plasma frame
:
\begin{equation}
u^0=\frac{u^\mu P_\mu}{\sqrt{s}}=\frac{E_{\bar{p}}+ E_{\bar{k}}}{\sqrt{s}}\,\qquad \quad
u^z =-\frac{u^\mu Q_\mu}{\sqrt{s}}=-\frac{(E_{\bar{p}}- E_{\bar{k}})}{\sqrt{s}}\,.
\end{equation}
The $u^x$ component can be determined from the condition $u^\mu u_\mu = 1$:
\begin{eqnarray}
u^x&=&\sqrt{u_0^2-u_z^2-1}=\frac{1}{\sqrt{s}}\sqrt{(E_{\bar{p}}+E_{\bar{k}})^2-(E_{\bar{p}}-E_{\bar{k}})^2-s}\nonumber\\
&=&\frac{1}{\sqrt{s}}\sqrt{4 E_{\bar{p}} E_{\bar{k}}-s}=\frac{1}{\sqrt{s}}\sqrt{2 E_{\bar{p}} E_{\bar{k}} (1+\cos\theta_{\bar{p}\bar{k}})}\,.
\end{eqnarray}

Putting everything together we find that $u^\mu p'_\mu$ is given by
\begin{equation}
\begin{split}
u^\mu p'_\mu & =u^0 \frac{\sqrt{s}}{2} - u^x\frac{\sqrt{s}}{2}\sin\theta\cos\phi - u^z\frac{\sqrt{s}}{2}\cos\theta\\
    & = \frac{E_{\bar{p}}+E_{\bar{k}}}{2}-\frac{1}{2}\sqrt{2 E_{\bar{p}} E_{\bar{k}}(1+\cos\theta_{\bar{p}\bar{k}})}\sin\theta\cos\phi + \frac{(E_{\bar{p}} - E_{\bar{k}})}{2}\cos\theta\\
    & = \frac{1}{2}\left(E_{\bar{p}} (1+\cos\theta) + E_{\bar{k}} (1-\cos\theta) - \sqrt{2 E_{\bar{p}} E_{\bar{k}}(1+\cos\theta_{\bar{p}\bar{k}})}\sin\theta\cos\phi\right)\,.
\end{split}
\end{equation}
Similarly we find
\begin{equation}
u^\mu k'_\mu = \frac{1}{2}\left(E_{\bar{p}} (1-\cos\theta) + E_{\bar{k}}(1+\cos\theta) + \sqrt{2E_{\bar{p}} E_{\bar{k}} (1+\cos\theta_{\bar{p}\bar{k}})}\sin\theta\cos\phi\right)\,.
\end{equation}
Finally, the Mandelstam variables are given by
\begin{equation}
t = -\frac{s}{2}(1-\cos\theta)\,\qquad\quad
s = 2E_{\bar{p}} E_{\bar{k}} (1-\cos\theta_{\bar{p}\bar{k}})\,.
\end{equation}

\subsubsection{Evaluation of the ${\cal K}_2$ kernel}

We now discuss the evaluation of the ${\cal K}_2$ kernel:
\begin{equation}
    {\cal K}_2=\frac{1}{(2\pi)^5}\int\frac{d^3{\bf k}\,d^3{\bf k'}}{2E_k\,2E_{k'}}|{\cal M}|^2f_0(u^\mu k_\mu)(1\pm f_0(u^\mu k'_\mu))\delta^4(p+k-p'-k')
\end{equation}
Also in this case we follow ref.~\cite{DeGroot:1980dk}. We introduce the four-vectors
\begin{equation}
\begin{split}
   &K^\mu = k^\mu+k'^\mu\\
   &P^\mu=p^\mu+p'^\mu\\
   &Q'^\mu=k^\mu-k'^\mu\\
   &Q^\mu=p^\mu-p'^\mu
\end{split}\,.
\end{equation}
Recalling that
\begin{equation}
\frac{d^3{\bf k}\,d^3{\bf k'}}{2E_k\, 2E_{k'}}=d^4k\,d^4k'\,\theta(E_k)\theta(E_{k'})\delta(k^2)\delta(k'^2)\,,
\end{equation}
we can use as integration variables $K$ and $Q'$ finding
\begin{equation}
\frac{d^3{\bf k}\,d^3{\bf k'}}{2E_k\,2E_{k'}}=\frac{1}{4}d^4K\,d^4Q'\,\theta(K_0)\theta(K^2)\delta(K^2+Q'^2)\delta(K^\mu Q'_\mu)\,.
\end{equation}
Since $\delta^4(p+k-p'-k')=\delta^4(Q+Q')$, we can integrate over $Q'$ obtaining
\begin{equation}
{\cal K}_2=\frac{1}{(2\pi)^5}\int \frac{1}{4}d^4K\theta(K_0)\delta(K^2+Q^2)\delta(K^\mu Q_\mu)|{\cal M}|^2 f_0(u^\mu k_\mu)(1\pm f_0(u^\mu k'_\mu))
\end{equation}
In the massless case
\begin{equation}
Q^2=-P^2=t\,,
\end{equation}
hence
\begin{equation}
\delta(K^2+Q^2)=\delta(K^2+t)\,.
\end{equation}
Using the identity
\begin{equation}
d^4K\,\theta(K^0)\delta(K^2+t)=\frac{d^3{\bf K}}{2\sqrt{{\bf K}^2-t}}\,,
\end{equation}
we can rewrite ${\cal K}_2$ as
\begin{equation}
{\cal K}_2=\frac{1}{8(2\pi)^5}\int\frac{d^3{\bf K}}{\sqrt{{\bf K}^2-t}}\delta(K^\mu Q_\mu)|{\cal M}|^2f_0(u^\mu k_\mu)(1\pm f_0(u^\mu k'_\mu))\,.
\end{equation}

We can now rewrite this formula in the COM frame, in which $P^\mu=(\sqrt{-t},0,0,0)$ and $Q^\mu=(0,0,0,\sqrt{-t})$.
Introducing polar coordinates for ${\bf K}$, 
with polar angles $\theta$ and $\phi$, one gets
\begin{equation}
\delta(K^\mu Q'_\mu)=\delta(|{\bf K}|\sqrt{-t}\cos\theta)=\frac{1}{|{\bf K}|\sqrt{-t}}\delta(\cos\theta)\,,
\end{equation}
which allows to trivially perform the integration over $\cos \theta$, leading to
\begin{equation}
{\cal K}_2=\frac{1}{8(2\pi)^5}\int\frac{|{\bf K}|\,d|{\bf K}|\,d\phi}{\sqrt{{\bf K}^2-t}\sqrt{-t}}|{\cal M}|^2 f_0(u^\mu k_\mu)(1\pm f_0(u^\mu k'_\mu))\,.
\end{equation}

As a last step, we need to determine the expressions for the $u^\mu$  components.
Focusing on $u^\mu k_\mu$ we find
\begin{equation}
u^\mu k_\mu =\frac{u^\mu}{2} (K_\mu+Q'_\mu)=\frac{u^\mu}{2}(K_\mu-Q_\mu)=\frac{1}{2}\left(u^0 \sqrt{{\bf K}^2-t}-u^x|{\bf K}|\cos\phi+u^z\sqrt{-t}\right)\,.
\end{equation}
Where we chose the orientation of the COM frame in such way that $u^y=0$. In an analogous way we find
\begin{equation}
u^\mu k'_\mu=\frac{u^\mu}{2} (K_\mu-Q'_\mu)=\frac{u^\mu}{2}(K_\mu+Q_\mu)=\frac{1}{2}\left(u^0 \sqrt{{\bf K}^2-t}-u^x|{\bf K}|\cos\phi-u^z\sqrt{-t}\right)\,.
\end{equation}
Exploiting the fact that $P^\mu/\sqrt{-t}$ and $Q^\mu/\sqrt{-t}$ coincide with the versors in the time and $z$ directions, we can write
\begin{equation}
\begin{split}
u^0&= \frac{u^\mu P_\mu}{\sqrt{-t}}=\frac{ E_{\bar{p}}+ E_{\bar{p}'}}{\sqrt{-t}}\,,\\
u^z&=-\frac{u^\mu Q_\mu}{\sqrt{-t}}=-\frac{E_{\bar{p}}- E_{\bar{p}'}}{\sqrt{-t}}\,.
\end{split}
\end{equation}
Finally, from $u^\mu u_\mu=1$, one gets
\begin{equation}
u^x=\sqrt{\frac{(E_{\bar{p}}+ E_{\bar{p}'})^2}{-t}-\frac{(E_{\bar{p}}- E_{\bar{p}'})^2}{-t}-1}=\frac{1}{\sqrt{-t}}\sqrt{2 E_{\bar{p}} E_{\bar{p}'}(1+\cos\theta_{\bar{p}\bar{p}'})}\,.
\end{equation}

In order to make the numerical evaluation of the kernel more stable, we used the following coordinate change $|{\bf K|}=\sqrt{-t}\tan\theta$, and then we defined $1/\cos\theta=x$. The expression for ${\cal K}_2$ becomes
\begin{equation}
{\cal K}_2 =\frac{1}{8(2\pi)^5}\int_1^\infty \int_0^{2\pi}dx\,d\phi\,|{\cal M}|^2f_0(u^\mu k_\mu)(1\pm f_0(u^\mu k'_\mu))
\end{equation}
with
\begin{equation}
\begin{split}
u^\mu k_\mu &=\frac{1}{2}\left((E_{\bar{p}}+E_{\bar{p}'})x-\sqrt{x^2-1}\sqrt{2E_{\bar{p}} E_{\bar{p}'}(1+\cos\theta_{\bar{p}\bar{p}'})}\cos\phi+(E_{\bar{p}}-E_{\bar{p}'})\right)\,,\\
u^\mu k_\mu &=\frac{1}{2}\left((E_{\bar{p}}+E_{\bar{p}'})x-\sqrt{x^2-1}\sqrt{2E_{\bar{p}} E_{\bar{p}'}(1+\cos\theta_{\bar{p}\bar{p}'})}\cos\phi-(E_{\bar{p}}-E_{\bar{p}'})\right)\,,\\
s &=\frac{-t}{2}(x+1)\,,\\
u &=\frac{t}{2}(x-1)\,.
\end{split}
\end{equation}

\providecommand{\href}[2]{#2}\begingroup\raggedright\endgroup

\end{document}